\documentclass[onepage,12pt,a4paper,pdftex]{article}

\usepackage{titletoc}
\usepackage{graphicx, amsfonts, amsmath, amssymb, amscd,  theorem, exscale,
epic, eepic, epsfig}

\usepackage{hyperref} %

\usepackage{makeidx}
\makeindex

\newcounter{Figure}
\theoremstyle{plain}
\theoremheaderfont{\scshape}

\newtheorem{Def}{\bf Definition}
\newtheorem{The}{\bf Theorem}

\newtheorem{Lem}{\bf Lemma}

\theorembodyfont{\upshape}

\theoremheaderfont{\scshape\small} \theorembodyfont{\scshape\small}

\newcommand{\real}{ {\mathbb R} }

\newcommand{\slap}{\mbox{$ \triangle \mkern -13mu / \ $}}
\newcommand{\nlap}{\mbox{$ \nabla \mkern -13mu / \ $}}
\newcommand{\dlap}{\mbox{$ div \mkern -13mu / \ $}}
\newcommand{\Dlap}{\mbox{$ D \mkern -13mu / \ $}}

\newcommand{\clap}{\mbox{$ curl \mkern -23mu / \ $}}

\newcommand{\Lbar}{\underline{L}}
\newcommand{\be}{\begin{equation}}
\newcommand{\ee}{\end{equation}}
\newcommand{\bea}{\begin{eqnarray}}
\newcommand{\eea}{\end{eqnarray}}
\newcommand{\beas}{\begin{eqnarray*}}
\newcommand{\eeas}{\end{eqnarray*}}

\newcommand{\Lie}{ {\mathcal L} }

\newcommand{\Nm}{\mathcal{N}}
\newcommand{\Lu}{\underline{L}}
\newcommand{\Td}{\mbox{$ \ ^{(T)} \pi $}}

\begin{document}

\begin{center}
{\Large \bf Neutrino Radiation Showing a Christodoulou \\  \vspace{2pt}  Memory Effect 
in General Relativity } \\ 
\end{center}
\vspace{5pt}
\begin{center}
{\large \bf Lydia Bieri \footnote{L. Bieri is supported by NSF grants DMS-1253149 and DMS-0904760. \\  
Lydia Bieri, University of Michigan, Department of Mathematics, Ann Arbor MI. lbieri@umich.edu}
, David Garfinkle \footnote{D. Garfinkle is supported by NSF grants PHY-0855532 and PHY-1205202. \\
David Garfinkle, Oakland University, Department of Physics, Rochester MI and University of Michigan, Michigan Center for Theoretical Physics, Randall Laboratory of Physics, Ann Arbor, MI. garfinkl@oakland.edu}} \\ 
\end{center}
{\bf Abstract:} 
We describe neutrino radiation in general relativity by introducing the energy-momentum tensor of a null fluid into the Einstein equations. Investigating the geometry and analysis at null infinity, we prove that a component of the null fluid enlarges the Christodoulou memory effect of gravitational waves. The description of neutrinos in general relativity
as a null fluid can be
regarded as a limiting case of a more general description using the massless limit of the Einstein-Vlasov system.  The present authors with co-authors have work in progress to generalize the results of this paper using this more general description. Gigantic neutrino bursts occur in our universe in core-collapse supernovae and in the mergers of neutron star binaries.

\section{Introduction and Main Results}
In this paper, we prove that there is a nonlinear memory effect for neutrino radiation. We describe the neutrinos from a typical source like binary neutron star merger or core-collapse supernova as a null fluid in the Einstein equations. We compute the radiated energy, derive limits at null infinity and compare them with
the Einstein vacuum (EV) case and the Einstein-Maxwell (EM) case.

We consider the Einstein equations 
\be \label{equsGRgRT}
G_{ij} \  := \ R_{ij} \ - \ \frac{1}{2} \ g_{ij} \ R \ = \ 8 \pi \
T_{ij}  \ , 
\ee
(setting $G = c  = 1$), $i, j = 0,1,2,3$.  
$g_{ij}$ denotes the metric tensor,  
$R_{ij}$ is the Ricci curvature tensor,  
$R$ the scalar curvature tensor, 
$G_{ij}$ denotes the Einstein tensor and 
$T_{ij}$ is the energy-momentum tensor. 

We describe the burst of neutrinos in a typical source such as core-collapse supernovae and binary neutron star mergers 
as a null fluid. This means that the energy-momentum tensor will have the form 
\be \label{nullfluid1}
T^{ij} = \mathcal{N} K^i K^j
\ee
with $K$ denoting a null vector to be specified below and $\mathcal{N}$ a positive scalar function. 
Later on, we make use of $\sqrt{\Nm} K^i = k^i$ and 
 \be \label{nullfluid2}
T^{ij} =  k^i k^j  \ . 
\ee
Let $L$ as well as $\underline{L}$ denote null vectors, properly defined below. 
Note that we refer to $L$ as the null vectorfield generating the corresponding outgoing null hypersurfaces and to $\underline{L}$ as the null vectorfield generating corresponding incoming null hypersurfaces. $L$ and $\underline{L}$ are complemented to a null frame by the $S$-tangent orthonormal frame $(e_A, A=1,2)$. 
Thus, the corresponding components of the energy-momentum tensor are 
\bea
& & T^{LL}, T^{\underline{L} \underline{L}} , 
 T^{L \underline{L}}, T^{AL} ,   T^{A\underline{L}} , 
  T^{AB}  \label{T1nullvector}
\eea

Initially, the burst will take off in all directions, but eventually the null part $T^{LL}$ will dominate, as we prove below.


%



Recall that the contravariant tensor $T^{ij}$ turns into the covariant tensor $T_{ij}$ by contracting with the metric. 
Thus we have 
\[
T^{LL} = \frac{1}{4} T_{\Lbar \Lbar} \ . 
\]

We show that the components of the energy-momentum tensor have the following decay behavior: 
\beas
T^{LL} & = & O(r^{-2} \tau_-^{-4}) \\ 
T^{AL} & = &  O(r^{-3} \tau_-^{-3}) \\ 
T^{AB} & = & O(r^{-4} \tau_-^{-2}) \\ 
T^{L \underline{L}} & = & O(r^{-4} \tau_-^{-2}) \\ 
T^{A \underline{L}} & = &  O(r^{-5} \tau_-^{-1}) \\ 
T^{\underline{L} \underline{L}} & = & O(r^{-6})  
\eeas
This means 
\beas
T_{\Lbar \Lbar} & = & O(r^{-2} \tau_-^{-4}) \\ 
T_{A \Lbar} & = &  O(r^{-3} \tau_-^{-3}) \\ 
T_{AB} & = & O(r^{-4} \tau_-^{-2}) \\ 
T_{L \underline{L}} & = & O(r^{-4} \tau_-^{-2}) \\ 
T_{A L} & = &  O(r^{-5} \tau_-^{-1}) \\ 
T_{LL} & = & O(r^{-6})  
\eeas
The quantity $\tau_-$ is defined as $\tau_- = \sqrt{1 + u^2}$. We use foliations of our spacetime by a time function $t$ and an optical function $u$ as explained below. 

We state our equations with all the components of the energy-momentum tensor. Their decay behavior emerges from the physical model and the corresponding mathematical consequences. 

In this paper, the covariant differentiation on the spacetime $M$ we denote by $D$ or $\nabla$, whereas the one on a spacelike hypersurface $H$ is 
$\overline{\nabla}$ or $\nabla$. It is clear from the context what $\nabla$ refers to.

The twice contracted Bianchi identities imply that 
\be \label{DG01}
D^{j} G_{ij} = 0 \ \ . 
\ee
This is equivalent to the following equation, namely, that the divergence of the
energy-momentum tensor of the null fluid vanishes: 
\be \label{DT01}
D^{j} T_{ij} = 0 . 
\ee

The Einstein equations (\ref{equsGRgRT}) for a null fluid reduce to 
\be \label{Einsteinnullfluid1}
R_{ij}  =   8 \pi T_{ij}  
\ee

We prove that there is a contribution from neutrino radiation to the nonlinear Christodoulou memory effect of gravitational waves. 
When describing neutrino radiation by a null fluid and coupling the energy-momentum tensor of a null fluid to the Einstein vacuum equations, 
we find the energy radiated away per unit angle in a given direction to be $F/8\pi$ with 
\be \label{nonlmn1}
F(\cdot) = \int_{- \infty}^{+ \infty} \left( \mid \Xi(u, \cdot) \mid^2 + 4 \pi T_{\Lbar \Lbar}^* (u, \cdot) \right) du  \ . 
\ee
The limit $T_{\Lbar \Lbar}^*$ of $T_{\Lbar \Lbar}$ is positive. See equation (\ref{nullfluid1}) and the subsequent paragraph. 

Considering gravitational radiation in the presence of an electromagnetic field as well as neutrino radiation, as present typically in binary neutron star mergers, we investigate 
the Einstein-Maxwell-null-fluid equations  and find similarly the energy radiated away per unit angle in a given direction to be $F/8\pi$ with 

\be \label{nonlmn2}
F(\cdot) = \int_{- \infty}^{+ \infty} \left( \mid \Xi(u, \cdot) \mid^2 +   \frac{1}{2} \mid A_F (u, \cdot) \mid^2  + 4 \pi T_{\Lbar \Lbar}^* (u, \cdot) \right) du  \ . 
\ee

The permanent displacement formula of the nonlinear memory effect involves formula (\ref{nonlmn1}) respectively (\ref{nonlmn2}). 
In both cases, we have a contribution from neutrino radiation given by the term $4 \pi  T_{\Lbar \Lbar}^* (u, \cdot) $. 

In a new project of the current authors together with P. Chen and S.-T. Yau, we imbed this result into a complete kinetic theory for neutrinos in general relativity. 

The Einstein-null-fluid equations (\ref{equsGRgRT}) respectively (\ref{Einsteinnullfluid1}) investigated in this article are proven to have 
geodesically complete solutions for physical initial data by the first present author in \cite{lydia5}. What could happen in general, is that the null geodesics may intersect. In order to prevent that, one has to specify corresponding initial data for the null fluid. We work with initial data which is asymptotically flat. Whereby the geometric part, that is the induced metric at time $t=0$ and the corresponding second fundamental form of the initial spacelike hypersurface may behave as in \cite{sta}. 
Necessary conditions on the null fluid for this spacetime to be non-singular are given in \cite{lydia5}. In particular, considering (\ref{nullfluid1}), 
in the region of compact support we choose the density to be $1$ and the vectorfield in the outside to decrease appropriately. Moreover, 
outside the region of compact support, the vectorfield is directed outwards. 
In \cite{lydia5} the Cauchy problem is solved to prove that the corresponding solution is non-singular. In particular, the local result is implied by the implicit function theorem, whereas the global result is achieved by a bootstrap argument.

\subsection{Nonlinear Christodoulou Memory Effect of Gravitational Waves}

The general theory of relativity predicts gravitational waves. There has been vast literature about this topic. Ongoing and future experiments like LIGO or LISA 
want to measure these waves directly. Other experiments based on radio astronomy aim at measuring the Christodoulou effect of these waves, which means that they would be detected through this nonlinear effect. 

A gravitational wave train will have two different effects on the test masses. A wave train traveling from its source to us, will pass the experiment. During the passage of such a wave train the test masses will experience `instantaneous displacements'. Afterwards, the test masses will show `permanent displacements'. The latter is the nonlinear memory effect (Christodoulou effect) of gravitational waves. Thus, the spacetime has been changed permanently. 
Such an effect was known in a linear theory \cite{zelpol}, but its contribution was very small and people believed it negligible. 
See also \cite{braggri}, \cite{bragthorne}. 
Christodoulou \cite{chrmemory} shows that 
this is a truly nonlinear effect and as such its contribution is much larger than expected. He computes and investigates exact solutions of the Einstein equations, no approximation is used. 

Since the pioneering days, when Christodoulou \cite{chrmemory} established his nonlinear result for the EV equations, 
showing that gravitational waves displace test masses permanently, 
it has been an open problem 
if electromagnetic fields in the EM equations contribute to the nonlinear effect. 
In the work of the first author with PoNing Chen and Shing-Tung Yau \cite{1lpst1}, \cite{1lpst2} we solve this problem and apply the new findings to astrophysical data. 
We show \cite{1lpst1} how the electromagnetic field in the EM equations contributes to the nonlinear memory effect of gravitational waves. And we investigate the effect on gravitational wave detectors. Precise formulas derived with geometric-analytic methods from the EM equations are related to experiment. And predictions for measurements are stipulated. We apply the new results to astrophysical data \cite{1lpst2}. 

We would like to emphasize that neither in the works \cite{1lpst1}, \cite{1lpst2} nor in this article, any approximation is used. The first author with Chen and Yau in 
\cite{1lpst1}, \cite{1lpst2} as well as the present authors in this paper derive and investigate exact solutions of the EM respectively Einstein-null-fluid equations. 
All the results mentioned in this subsection hold for large data. Examples of the latter include supernovae, mergers of black holes or neutron stars. 

Typical sources of neutrino radiation are big events in the universe such as core-collapse supernovae or binary neutron star mergers.  In a core collapse supernova, the 1.4 solar mass iron core of a massive star is converted into neutrons and neutrinos.  In the process, about $3 \times {{10}^{53}}$ ergs of gravitational binding energy is released, almost all of
it in the form of neutrino radiation.\cite{ott1}  In addition to the gravitational wave effects of the neutrino radiation,
the neutrinos themselves can be detected thus enhancing the possibility of detection through a joint gravitational wave and
neutrino search.\cite{ottetal}  Since neutrinos are the dominant form of energy loss, one 
might think that the gravitational waves generated by neutrinos (and the associated gravitational wave memory) would be
the most easily detected gravitational wave signature from core-collapse supernovae.  Unfortunately, this turns out not to
be the case.\cite{ott1,janka}  To begin with, the spherically symmetric part of the energy emission produces no gravitational
waves, so it is only the fraction (about 2\% or so\cite{janka}) of the neutrino 
energy emission that is anisotropic that produces 
gravitational waves.  These gravitational waves are higher amplitude than those produced by the matter in the supernova core;
however, their frequency is also smaller.  This is because though the neutrinos are produced promptly in the collapse of the 
core, it takes them a time of on the order of a second to escape from the extremely high density region of the collapsing core.  Thus neutrino emission and the associated gravitational waves have a time scale of about a second, or equivalently 
a frequency scale of about 1 Hz.  For LIGO and the other ground based gravity wave detectors, 1 Hz is the low frequency range
where seismic noise in the detectors is large.\cite{ligo}  This makes detecting gravitational wave memory in supernova core
collapse very challenging.  However, current improvements being made to gravity wave detectors include an improvement in seismic isolation and thus a lessening in detector noise at low frequency.  Thus in addition to the overall improvement in 
the possibility of detecting gravitational waves, the detectors should also have an improved possibility of detecting gravitational wave memory.  

A binary neutron star system consists of two neutron stars in orbit around each other.  Such a system loses energy in gravitational radiation, causing the neutron stars to orbit at ever smaller distances (referred to as ``inspiral'') 
until they merge.\cite{taylor}  
The merged object is a supermassive neutron star with large thermal energy as well as a flattened 
shape caused by its high rotation speed.  It is estimated\cite{dessart} that the supermassive neutron star radiates
neutrinos with a power of about $5 \times {{10}^{52}}$ erg/s over a time of about 1 s.  Eventually the supermassive 
neutron star is expected to collapse to form a black hole and perhaps generate a gamma ray burst.  The gravitational 
radiation generated by the inspiral of the neutron stars is considered to be the most promising candidate for detection of
gravitational waves.\cite{riles}  
The gravitational wave memory from the burst of neutrinos will be more challenging to detect because
of the longer time scale and the seismic noise in the detectors.

\section{Setting}

It will be useful to split the Riemannian curvature tensor $R_{\alpha \beta \gamma \delta}$ into its Weyl tensor $W_{\alpha \beta \gamma \delta }$ being traceless, and a part including the spacetime
Ricci curvature $R_{\alpha \beta}$ 
and spacetime scalar curvature $R$: 
\bea
R_{\alpha \beta \gamma \delta } &=&W_{\alpha \beta \gamma
\delta }+\frac{1}{2}(g_{\alpha \gamma }R_{\beta \delta }+
g_{\beta \delta }R_{\alpha \gamma }-g_{\beta
\gamma }R_{\alpha \delta }-g_{\alpha \delta }R_{\beta \gamma })  \nonumber \\
&&-\frac{1}{6}\left( g_{\alpha \gamma }g_{\beta \delta }- 
g_{\alpha \delta }g_{\beta \gamma }\right) R  \ \ . \label{Riem*tracelessRicci}
\eea

We work with two different foliations of the spacetime $(M, g)$. 
First, we can choose an appropriate time function $t$ and obtain a foliation given by the level sets $H_t$. We denote the time vector field by $T$, i.e. the 
future-directed normal to the foliation. Thus it is $T^i =  - \Phi^2 g^{ij} \partial_j t$ and $Tt = T^i \partial_j t = 1$. 
The resulting spacetime foliation is diffeomorphic to the product $\real \times \bar{M}$ with $\bar{M}$ being a $3$-manifold and 
each level hypersurface $H_t$ of $t$ is diffeomorphic to $\bar{M}$. 
The metric then reads 
\be
g = - \Phi^2  dt^2 + \bar{g}
\ee
where $\bar{g} = \bar{g} (t)$ is the induced metric on $H_t$. 
We denote the components of the inverse metric as $g^{ij} = (g^{-1})^{ij}$. 
The lapse function $\Phi$ is given as  $ \Phi := (- g^{ij} \partial_i t \partial_j t)^{- \frac{1}{2}}$. 
Choosing a frame field $\{ e_i \}$ for $i = 1, 2, 3$ on $H_t$, we Lie-transport it along the integral curves of $T$. Thus, it is 
\[
[T, e_i] = 0  \ . 
\]
Write $\bar{g}_{ij} = \bar{g}(e_i, e_j) = g(e_i, e_j)$. 
Then the second fundamental form $k$ is given by 
\bea
k_{ij} & = & k(e_i, e_j) \\ 
& = & \frac{1}{2} \Phi^{-1} \frac{\partial \bar{g}_{ij}}{\partial t} \ \ . 
\eea
We choose to work with a maximal time function, that is 
\be
tr k = 0 \ \ . 
\ee
Moreover, denote by $N$ the spacelike unit normal vector of $S_{t,u}$ in $H_t$ and from now on let $T$ be the timelike unit normal vector of $H_t$ in the spacetime. Often we shall use the frame $(T, N, e_2, e_1)$. 

Second, we work with a null foliation of the spacetime. 
For this purpose, we foliate the spacetime by an optical function $u$ and denote its lapse function by $a$. 
Now, the level sets $C_u$ of $u$ are outgoing null hypersurfaces. Along $C_u$ we pick a suitable pair of normal vectors. 
Denote by $e_4$ and $e_3$ the 
null pair, i.e. $g(e_3, e_4) = - 2$, where $e_4 = T+N$ and $e_3 = T-N$. 
We consider the intersection $S_{t,u} = H_t \cap C_u$. 
Let $\{ e_A \}, A = 1,2$ be an orthonormal frame on $S_{t,u}$. This yields a null frame $( e_4, e_3, e_2, e_1)$ in the spacetime. 
Often we write $e_4 = L$ and $e_3 = \underline{L}$. 

Thus, $\underline{L}$ is transversal to $C$, the latter being generated by $L$. 
This vectorfield  $L$ corresponds through the spacetime metric $g$ to the 
1-form $-du$. In an arbitrary frame, we write 
\be \label{Luarbframe1}
L^{\mu} \ = \ - g^{\mu \nu} \ \partial_{\nu} \ u \ . 
\ee

We also need the second fundamental forms with respect to $e_4$ and $e_3$ respectively. Let $X, Y$ be arbitrary tangent vectors to $S$ at a point in $S$. Then the second fundamental forms are defined to be 
\[
\chi (X, Y) = g(\nabla_X e_4, Y) \ \ , \ \ \underline{\chi} (X, Y) =  g(\nabla_X e_3, Y) \ \ . 
\]
Further, denote their traceless parts by $\hat{\chi}$ and $\hat{\underline{\chi}}$ respectively. These are in fact the shears. 

More details about these foliations can be found in \cite{sta} as well as in \cite{zip}, \cite{zip2} and \cite{lydia1}, \cite{lydia2}. 

Given this null pair, $e_3$ and $e_4$, we
can define the tensor 
of projection from the tangent space of $M$ to that of $S_{t,u}$.
\[  \Pi^{\mu \nu} = g^{\mu \nu} + 
\frac{1}{2}(e_4^{\nu}e_3^{\mu}+e_3^{\nu}e_4^{\mu}). \] 

We shall decompose the Einstein equations as well as the curvature and all the geometric quantities with respect to these two foliations. 
These decompositions were first introduced in \cite{sta} and then applied and further investigated in \cite{zip}, \cite{zip2} and in \cite{lydia1}, \cite{lydia2}. 
We refer to these works for the detailed procedures. 

Let $\underline{u} = u + 2r$ and $\tau_-^2 = 1 + u^2$ as well as $\tau_+^2 = 1 + \underline{u}^2$.

The following vectorfields are expressed in terms of $L$ and $\underline{L}$. 
The time vectofield $T$ reads 
\be \label{intT12}
T \ = \ \frac{1}{2} \ \big( L \ + \ \underline{L} \big)  \ \ . 
\ee
The generator $S$ of scalings is defined to be 
\be \label{intS12}
S \ = \ \frac{1}{2} \ \big( \underline{u} \ L \ + \ u \ \underline{L}  \big) \ \ . 
\ee
The generator $K$ of inverted time translations is defined as 
\be \label{intK12}
K \ = \ \frac{1}{2} \ \big( \underline{u}^2 \ L \ + \ u^2 \ \underline{L}  \big) \ \ . 
\ee
Then the vectorfield $\bar{K} = K + T$ is 
\be \label{intKbar12}
\bar{K} \ = \ \frac{1}{2} \ \big( \tau_+^2 \ L \ + \ \tau_-^2 \ \underline{L}  \big) \ \ . 
\ee

We decompose the second fundamental form $k_{ij}$ of $H_t$ according to 
\bea
k_{NN} &=& \delta \\
k_{AN} &=& \epsilon_A\\
k_{AB} &=& \eta_{AB}  \ \ . 
\eea
Then, define
\be \label{theta1}
  \theta_{AB} =  \langle \nabla_{A} N, e_B \rangle.
\ee   
More generally, the second fundamental form $\theta_{ab}$ with $a, b = \{ 1,  2, 3 \}$  of the $u$-foliation within $H$ is 
obtained by projecting $\nabla_s N_t$ from $H$ to $S$. Thus, the resulting tensor is tangent to $S$. 
Choosing on $H$ the orthonormal frame $\{ N, \{ e_A \}_{A = 1,2}  \}$ where $\{ e_1, e_2 \}$ is an orthonormal frame on $S$, 
we find formula (\ref{theta1}). It is then easy to see that relative to arbitrary coordinates on $S$, the second fundamental form reads 
\be \label{theta2}
\theta_{AB} = \frac{1}{2a} \frac{\partial}{\partial u} \gamma_{AB}
\ee
where $\gamma_{AB}$ denotes the induced metric on $S$.

The Ricci coefficients of the null standard frame $T-N,T+N,e_2, e_1$ are 
\bea
\chi^{\prime}_{AB} &=& \theta_{AB} - \eta_{AB} \\
\underline{\chi}^{\prime}_{AB} &=& - \theta_{AB} - \eta_{AB} \\
\underline{\xi}^{\prime}_A &= &\phi^{-1} \nlap_A \phi - a^{-1} \nlap_A a \\
\underline{\zeta}^{\prime}_A &= &\phi^{-1} \nlap_A \phi - \epsilon_A  \\
\zeta^{\prime}_A &= &\phi^{-1} \nlap_A \phi + \epsilon_A  \\
\nu^{\prime} &= & - \phi^{-1} \nlap_N \phi + \delta  \\
\underline{\nu}^{\prime} &= &  \phi^{-1} \nlap_N \phi + \delta   \\
\omega^{\prime} &=&\delta -a^{-1} \nlap_N a
\eea
The  Ricci coefficients of the null frame
$a^{-1}(T-N),a(T+N),e_2, e_1$ are denoted by $\chi$, $\underline{\chi}$, etc. 
In what follows, we drop the primes, but point out, which frame is used. 
\begin{Def}
Define the null components of the Weyl curvature tensor $W$ to be 
\bea
\underline{\alpha}_{\mu \nu} \ (W) \ & = & \ 
\Pi_{\mu}^{\ \rho} \ \Pi_{\nu}^{\ \sigma} \ W_{\rho \gamma \sigma \delta} \
e_3^{\gamma} \ e_3^{\delta} 
\label{underlinealpha} \\ 
\underline{\beta}_{\mu} \ (W) \ & = & \ 
\frac{1}{2} \ \Pi_{\mu}^{\ \rho} \ W_{\rho \sigma \gamma \delta} \  e_3^{\sigma} \
e_3^{\gamma} \ e_4^{\delta} 
\label{underlinebeta} \\ 
\rho \ (W) \ & = & \ 
\frac{1}{4} \ W_{\alpha \beta \gamma \delta} \ e_3^{\alpha} \ e_4^{\beta} \
e_3^{\gamma} \ e_4^{\delta} 
\label{rho} \\ 
\sigma \ (W) \ & = & \ 
\frac{1}{4} \ \ ^*W_{\alpha \beta \gamma \delta} \ e_3^{\alpha} \ e_4^{\beta} \
e_3^{\gamma} \ e_4^{\delta} 
\label{sigma} \\ 
\beta_{\mu}  \ (W) \ & = & \  
\frac{1}{2} \ \Pi_{\mu}^{\ \rho} \ W_{\rho \sigma \gamma \delta} \ e_4^{\sigma} \
e_3^{\gamma} \ e_4^{\delta} 
\label{beta} \\ 
\alpha_{\mu \nu} \ (W) \ & = & \ 
\Pi_{\mu}^{\ \rho} \ \Pi_{\nu}^{\ \sigma} \ W_{\rho \gamma \sigma \delta} \
e_4^{\gamma} \ e_4^{\delta}  \ . 
\label{alphaR}
\eea
\end{Def}
In \cite{sta} the following behavior is shown. 
\beas 
\underline{\alpha}(W) \ & = & \ O \ (r^{- 1} \ \tau_-^{- \frac{5}{2}}) \\ 
\underline{\beta}(W) \ & = & \ O \ (r^{- 2} \ \tau_-^{- \frac{3}{2}}) \\ 
\rho(W) \   & = & \ O \ (r^{-3})  \\ 
\sigma(W)  \ & = & \ O \ (r^{-3} \ \tau_-^{- \frac{1}{2}}) \\ 
\alpha(W) , \ \beta(W) \ & = & \ o \ ( r^{- \frac{7}{2}})  \ \ .
\eeas
\[\]

\subsection{Ricci Rotation Coefficients}

The Ricci rotation coefficients of the null frame are: 
\beas
\chi_{AB} & = & g(D_A e_4, e_B)  \\ 
\underline{\chi}_{AB} & = & g(D_A e_3, e_B)  \\ 
\underline{\xi}_A & = & \frac{1}{2} g(D_3 e_3, e_A)  \\ 
\zeta_A & = & \frac{1}{2} g(D_3 e_4, e_A)  \\ 
\underline{\zeta}_A & = & \frac{1}{2} g(D_4 e_3, e_A)  \\ 
\nu & = & \frac{1}{2} g(D_4 e_4, e_3)  \\ 
\underline{\nu} & = & \frac{1}{2} g(D_3 e_3, e_4)  \\ 
\epsilon_A & = & \frac{1}{2} g(D_A e_4, e_3)
\eeas
In \cite{sta} the authors compute fundamental derivatives, of which here we use: 
\beas
D_4 e_A & = & \Dlap_4 e_A + \underline{\zeta}_A e_4 \\ 
D_4 e_3 & = & 2 \underline{\zeta}_A e_A + \nu e_3  \ . 
\eeas

The notation for $\zeta$ and $\underline{\zeta}$ above as introduced in \cite{sta} is used slightly in a different way in this paper. We explain the underlying structures next. 
In order to do so, we introduce the {\itshape torsion $1$-form} $\zeta$ on $S$ as 
\begin{equation}
\zeta (X) =  \frac{1}{2}  g  (\nabla_XL,  \underbar{L}) \ \ \ \forall \ X \ \in \ TS \ . 
\end{equation}

One can then show that 
\begin{equation}
\nabla_L \underline{L} \ = \ - \ 2 \ Z \ 
\end{equation}
is the $S_s$-tangent vectorfield corresponding to the 1-form $\zeta$. 
We recall $g(L, \underline{L}) = -2$ and $\nabla_L L = 0$, $\nabla_{\underline{L}} \underline{L} = 0$. 
Then it is 
\[
L g(L, \Lbar) = g(\underbrace{\nabla_L L}_{= 0}, \Lbar) + g (L, \underbrace{\nabla_L \Lbar}_{= - 2 Z}) = 0 . 
\]
Thus we have 
\[
g(L, \nabla_L \Lbar) = 0 
\]
telling us that $\nabla_L \Lbar$ is tangential to $C$. 
Moreover, compute 
\[
L g(\Lbar, \Lbar) =  2 g (\nabla_L \Lbar, \Lbar) = 0 
\]
yielding 
\[
g(\Lbar, \nabla_L \Lbar) = 0 . 
\]
Therefore, $\nabla_L \Lbar$ does not have any component along $L$ either but is indeed tangential to $S$. This fact is used in (\ref{coeff*1}).

Let us further explore the torsion. 
Take any $X \in  T_xS$ and extend $X$ to a 
Jacobi field along the generator through $x$. We compute 
\begin{eqnarray*}
g \ (\nabla_L \underline{L}, \ X) \ & = & \ 
L \ ( \ \underbrace{g \ (\underline{L}, \ X)}_{=0}) \ - \ 
g \ (\underline{L}, \nabla_L X) \\ 
\ & = & \ - \ g  \ (\underline{L}, \ \nabla_XL)  \\ 
\ & = & \ - \ 2 \ \zeta \ (X) \ . 
\end{eqnarray*}
The second line holds because $[L,  X]  =  0$. 
Then consider any vector $X  \in  T_xS$ together with 
$g(\nabla_{\underline{L}}L, X)$. 
To calculate the following, we use formula (\ref{Luarbframe1}) in an arbitrary frame:  
\[
L^{\mu} \ = \ - g^{\mu \nu} \ \partial_{\nu} \ u \ . 
\]
We make use of the fact that the Hessian of a function is symmetric and we compute 
\begin{eqnarray*}
g \ (\nabla_{\underline{L}}L, \ X) \ & = & \ - \ \nabla^2 u \ \cdot \ 
(\underline{L}, \ X ) \\ 
\ & = & \  - \ \nabla^2 u \ \cdot \ (X, \ \underline{L}) , \\ 
\ & = & \ g \ ( \nabla_XL, \ \underline{L}) \\ 
\ & = & \ 2 \ \zeta \ (X) \ . 
\end{eqnarray*}
Summarizing, we find that 
$\zeta_A =  - \frac{1}{2} g(\nabla_L \underline{L}, e_A) = \frac{1}{2} g(\nabla_{\underline{L}} L, e_A)$. 
Thus, in the notation from above, this means $\zeta_A = - \underline{\zeta}_A$. 

Then the corresponding derivatives take the form: 
\bea
D_4 e_A & = & \Dlap_4 e_A - \zeta_A e_4  \label{derR*1} \\ 
D_4 e_3 & = & - 2 \zeta_A e_A + \nu e_3  \ .   \label{derR*2}
\eea

Further, we calculate 
\[
g(\nabla_L e_A , \Lbar) = \underbrace{L g(e_A, \Lbar)}_{=0} \underbrace{ - g(e_A, \nabla_L \Lbar) }_{= 2 \zeta_A} = 2 \zeta_A . 
\]
In a similar way, it is shown that 
$g(\nabla_L e_A, L) = 0$ and $g(\nabla_L e_A, e_B) = 0$.

\subsection{Behavior of the Energy-Momentum Tensor}

{\bf Behavior and decay of vector fields at null infinity.} 
For the current purpose, let us use the notation as in equation (\ref{nullfluid1}), where we read 
\[
T^{ij} = \mathcal{N} K^i K^j \ \ . 
\]
The vector $k$ initially will be of the form 
\be \label{k}
k = a L + b \underline{L} + V 
\ee
with $V$ denoting a vector tangent to $S$. 

In the following we show that a long time after the burst the $L$ direction will dominate, that is that $\underline{L}$ and $V$ will decay along $L$. 
Moreover, we show that $b$ decays and that $a$ converges to $1$ in the corresponding limits.

Let us first consider the vectorfield $L$ and write 
\[
T^{ij} = \Nm L^i L^j \ . 
\]
We have seen that the twice contracted Bianchi identities imply (\ref{DG01}), and therefore the Einstein equations enforce (\ref{DT01}), that is 
\[
\nabla_i T^{ij} = 0 \ . 
\]
Thus we have 
\beas
(\nabla_i \Nm) L^i L^j = - \Nm (\nabla_i L^i) L^j - \Nm L^i (\nabla_i L^j) \\ 
(\nabla_L \Nm) L^j = - \Nm div L L^j - \Nm \nabla_L L^j  \ . 
\eeas
As $L$ is a geodesic vectorfield, the last term is zero and we have  
\be \label{nablaLN1}
(\nabla_L \Nm) = - \Nm div L \ . 
\ee

From the studies in \cite{sta} we know how the geodesics behave. Some of this geometric structure worked out in \cite{sta} is used in the following. 
In particular, it is 
\[
\lim_{C_u, t \to \infty} r tr \chi = 2 \ \ , \ \ \ \lim_{C_u, t \to \infty} r tr \underline{\chi} = - 2  \ . 
\]
We compute 
\bea
div L & = &  tr \chi + l.o.t.  = \frac{2}{r} + l.o.t.  \\ 
div \underline{L} & = & tr \underline{\chi} + l.o.t.  = - \frac{2}{r} + l.o.t.
\eea
Then we have in equation (\ref{nablaLN1}) 
\[
(\nabla_L \Nm) = - \Nm div L = - 2 \Nm r^{-1} + l.o.t. 
\]
Therefore, we conclude that 
\[
\Nm = O(r^{-2}) \ . 
\]
Thus we have 
\[
T^{LL}  = O(r^{-2}) \ . 
\]
We can absorb $\Nm$ into $L$ and denote $\sqrt{\Nm} L^i = L'^i$. 
However, we continue by dropping the prime and will point out which vectorfield is used. 
We use the following two settings: 1) we take the vectorfield $L$ to generate an affinely parametrized geodesic, and 
2) $L$ stands for $L'$. 

Working with the geodesic vectorfield $L$ we compute decay rates for $e_A$ and $\Lu$ along $L$. 
Next, we focus on the vectorfield $e_A$. 
We find from (\ref{derR*1}) that (\ref{coeff*2}) holds and by direct computation (\ref{coeff*3}) holds for a vector $V$ tangential to $S$: 
\bea
\nabla_L e_A & = & - \zeta_A L  \label{coeff*2}  \\ 
\nabla_L V & = & \chi^A_{ \ B} V^B e_A - V^A \zeta_A L .   \label{coeff*3}
\eea

In order to find the behavior of $e_A$ along $L$, focus on equation (\ref{coeff*2}) and use the fact that $\zeta = O(r^{-2})$  
to find 
\[
\nabla_L e_A  =  - \zeta_A L = O(r^{-2})   \ . 
\]
Thus, along $L$ the vectorfield $e_A$ decays like $r^{-1}$.

Finally, consider vectorfield $\underline{L}$. Equation (\ref{derR*2}) exhibits a component along $\underline{L}$. But we show above that 
this component vanishes and $\nabla_L \underline{L}$ is tangential to $S$. 
Now, with $\zeta = O(r^{-2})$ 
we find  
\bea
\nabla_L \underline{L} & = & - 2 \zeta_A e_A   =  O(r^{-2}) O(r^{-1}) = O(r^{-3}) \ .   \label{coeff*1}  \\ 
\eea
And along $L$ the vectorfield $\underline{L}$ decays like $r^{-2}$. 

Switching to the vectorfield $L = L'$, we point out: 
Consequently, we find that along $L = L'$ the vectorfield $\Lu$ decays like $r^{-3}$ and for $A = \{ 1, 2 \}$ the vectorfield $e_A$ decays like $r^{-2}$. 

It then directly follows that {\bf the energy-momentum tensor has the following behavior in $\mathbf{r}$}: 
\beas
T^{LL}  & = & O(r^{-2})  \\ 
T^{AL} & = &  O(r^{-3}) \\ 
T^{AB} & = & O(r^{-4} ) \\ 
T^{L \underline{L}} & = & O(r^{-4}) \\ 
T^{A \underline{L}} & = &  O(r^{-5} ) \\ 
T^{\underline{L} \underline{L}} & = & O(r^{-6})  \ . 
\eeas

{\bf Remark:} 
When absorbing $\mathcal{N}$ into $L$ denoting $L' = \sqrt{\mathcal{N}} L$, 
then the components of the energy-momentum tensor take the previous form. In the discussion to follow we take the vectorfield $L$ to generate an affinely parametrized geodesic. Thus, the following holds: 
along $L$ the vectorfield $\Lu$ decays like $r^{-2}$ and for $A = \{ 1, 2 \}$ the vectorfield $e_A$ decays like $r^{-1}$.

The null vector $k$ satisfies 
\[
k_a k^a = 0  . 
\]
Thus, we have 
\[
0 = (a L_a + b  \underline{L}_a + V_a) (a L^a + b \underline{L}^a + V^a)  
= -4 ab + V_a V^a . 
\]
That is 
\be \label{4abVV}
4ab =  V_a V^a . 
\ee


The goal is to prove that $k$ tends to $L$ along $C_u$ as $t \to \infty$. 

For that purpose, we first investigate the deformation tensor for $T$. Let $Y, Z$ denote any vectorfields in $M$. 
The deformation tensor of $T$ then is 
\be
(\Lie_T g ) (Y, Z) = g(\nabla_Y T, Z) + g(Y, \nabla_Z T) \ . 
\ee
We denote the components of the deformation tensor for $T$ by 
\be
\ ^{(T)} \pi_{\alpha \beta}  = (\Lie_T g )_{\alpha \beta} . 
\ee

With (\ref{intT12}) we write 
\bea
(\Lie_T g ) (Y, Z) & = & \frac{1}{2} (\Lie_L g) (Y, Z) + \frac{1}{2} (\Lie_{\Lu} g) (Y, Z) \\ 
& = & \frac{1}{2} g(\nabla_Y L, Z) + \frac{1}{2} g(Y, \nabla_Z L) 
+ \frac{1}{2} g(\nabla_Y \Lu, Z) + \frac{1}{2} g(Y, \nabla_Z \Lu)  \label{defTL1}  
\eea

In a null frame, the vectorfields $Y$ and $Z$ read 
$Y = Y^L L + Y^{\Lu} \Lu + Y^A e_A$ and \\ 
$Z = Z^L L + Z^{\Lu} \Lu + Z^A e_A$ with $A = \{ 1,2 \}$. Then, with respect to a general null frame the deformation tensor of $T$ decomposes into the following components: 
\bea
(\Lie_T g ) (Y, Z) & = & 
Y^L Z^L   \Td_{44} + (Y^L Z^{\Lu} + Y^{\Lu} Z^L) \Td_{34} + (Y^A Z^L + Y^L Z^A) \Td_{A4} \nonumber \\ 
& & + Y^{\Lu} Z^{\Lu} \Td_{33} + (Y^A Z^{\Lu} + Y^{\Lu} Z^A) \Td_{3A} + Y^A Y^B \Td_{AB}  \ . 
\eea
Direct computations yield 
\beas
\Td_{44} & = & - 2 \nu \\ 
\Td_{34} & = & 2 \delta \\ 
\Td_{A4} & = & - 2 \epsilon_A + \phi^{-1} \nlap_A \phi \\ 
\Td_{33} & = & - 2 \underline{\nu} \\ 
\Td_{3A} & = & 2 \epsilon_A + \phi^{-1} \nlap_A \phi \\ 
\Td_{AB} & = & - 2 \eta_{AB}  
\eeas
with the right hand sides being the connection coefficients introduced earlier. Note that for geodesic vectorfields some of these terms vanish.

In the case where $Y$ and $Z$ are tangential to $S$, equation (\ref{defTL1}) reduces to 
\bea
(\Lie_T g ) (Y, Z)  & = & 
\chi (Y, Z) + \underline{\chi} (Y, Z) = -2 \eta (Y, Z) \label{defTL2} \\ 
& = & 
tr \chi + \hat{\chi} + tr \underline{\chi} + \hat{\underline{\chi}} \\ 
& = & 
\hat{\chi} + \underline{\hat{\chi}} + O(r^{-2}) \ . 
\eea
The last equation holds because of 
\[
tr \chi = \frac{2}{r} +  O(r^{-2})  \ \ \ tr \underline{\chi}  = - \frac{2}{r} +  O(r^{-2})  \ . 
\]
Further, as $\hat{\underline{\chi}}_{AB} = O(r^{-1})$ is the lowest order term in (\ref{defTL1}) respectively in (\ref{defTL2}), 
we find that the deformation tensor for $T$ behaves like 
\[
(\Lie_T g ) (Y, Z) \gamma_{AB} =   Y^A Z_A \hat{\underline{\chi}}_{AB} + l.o.t.  \ . 
\]

Moreover, we recall the facts that for any $Y, Z \in T_xC$ it is 
$\chi (Y, Z) = \chi(\Pi^C Y, \Pi^C Z)$ and for any $Y, Z \in T_x \underline{C}$ it is 
$\underline{\chi}(Y,Z) = \underline{\chi} (\Pi^{\underline{C}} Y, \Pi^{\underline{C}} Z)$ with $\Pi^C$ denoting the projection along $L$ onto the tangent space of $S$ and 
$\Pi^{\underline{C}}$ the projection along $\Lu$ onto the tangent space of $S$. 

Next, we need to get bounds on the coefficients $V^A$ of the above vectorfield $V = V^A e_A$. 
For this purpose, we consider the deformation tensor of the vectorfield $K$ introduced in (\ref{intK12}). 
With $V$ being tangential to $S$ we find 
\bea
(\Lie_K g ) (V, V)  & = & \frac{1}{2} \underline{u}^2 (\Lie_L g) (V, V) + \frac{1}{2} u^2 (\Lie_{\Lu} g) (V, V) \nonumber \\ 
& = & V^A V^B \underline{u}^2 \chi_{AB} + V^A V^B u^2 \underline{\chi}_{AB}  \\ 
& = &  V^A V^B \big{\{} \underline{u}^2 \hat{\chi}_{AB} +  u^2 \underline{\hat{\chi}}_{AB} 
+ \frac{1}{2} \underline{u}^2 tr \chi \gamma_{AB} + \frac{1}{2} u^2 tr \underline{\chi} \gamma_{AB}
\big{\}} \ . 
\eea
We split $(\Lie_K g )$ into its trace and traceless part, the latter being denoted as $\widehat{(\Lie_K g )}$. This gives 
\be
\widehat{(\Lie_K g )} (V, V) = V^A V^B \underline{u}^2 \hat{\chi}_{AB} + V^A V^B u^2 \underline{\hat{\chi}}_{AB}   \ . 
\ee
Observing the orders of these terms, we find
\[
\underline{u}^2 \hat{\chi}_{AB} = C_2 + O(r^{-1})  \mbox{  and  }
  u^2 \underline{\hat{\chi}}_{AB} = O(u^2 r^{-1}) \ . 
\]
As $\widehat{(\Lie_K g )}(V, V)$ has to go to zero as $t \to \infty$, we obtain bounds on the components $V^A$. 
We conclude for fixed $u$ and as $r \to \infty$ 
\bea
(\Lie_K g ) (V, V) \gamma_{AB} & = & 
 V^A V_A  \cdot   C_2 + l.o.t.  = C_2 | V |^2  + l.o.t.   \label{Vest*1}
\eea
As $t \to \infty$ the quantity $\widehat{(\Lie_K g )} (V, V)$ goes to zero.  
Therefore the coefficients $V^A$ have to decay like $r^{- \epsilon}$. 
Equation (\ref{Vest*1}) yields the bounds for the coefficients 
\[
V^A = O(r^{- \epsilon})  \ . 
\]

{\bf Proof of Convergence of $\mathbf{k}$ to $\mathbf{L}$ as $\mathbf{t} \mathbf{\to} \mathbf{\infty}$.} 
Let $k$ be a null geodesic, that is, 
\beas
k^a \nabla_a k =  \nabla_k k = 0 \\ 
k^a k_a = 0 \ . 
\eeas
If our manifold were the Minkowski space, then there would exist conformal Killing fields $X$, that is, 
\[
\nabla_{(a} X_{b)} = \phi g_{ab} 
\]
for some scalar $\phi$, that is 
\[
(\Lie_X g) = \phi g \ . 
\]
Then it follows that 
\[
k^a \nabla_a (X_a k^a) = 0 
\]
and consequently that  for each geodesic there exists a constant $c$ such that $k^a X_a = c$. 

As our manifold is a Lorentzian manifold with a lot of curvature structure, there are no `pure' conformal Killing fields, and the afore-mentioned equations in Minkowski space do not hold in a general Lorentzian manifold. However, another property of our Lorentzian manifold, namely its asymptotic flatness, guarantees the existence of almost- and quasi-conformal Killing fields. This means that the corresponding deformation tensors are suitably small and tend to zero as $t  \to \infty$ in a suitable way. 
In that case, the afore-mentioned equations `hold in an asymptotic sense'. 
This is what we have to prove now.

The null geodesic vectorfield $k$ takes the form as in (\ref{k})
\[
k = a L + b \underline{L} + V  \ . 
\]

Consider $(T_b k^b)$ and write in view of the above equations 
\be \label{nablakT*1}
\nabla_k (T_b k^b) = (\nabla_k T_b) k^b = (\nabla_k T) \cdot k = (\nabla_k T) \cdot (aL + b \Lu + V)
\ee

First, we investigate $\nabla_k T$. 
\beas
\nabla_k T & = & 
\frac{1}{2} a \nabla_L L + \frac{1}{2} b \nabla_{\Lu} L 
+  \frac{1}{2} V^A \nabla_A L 
+ \frac{1}{2} a \nabla_L \Lu 
+ \frac{1}{2} b \nabla_{\Lu} \Lu 
+ \frac{1}{2} V^A \nabla_A \Lu \ . 
\eeas
To compute the subsequent orders of the terms, we use the information from \cite{sta}.  
First, 
we derive 
\beas
\nabla_k T & = & 
0 \\ 
& & 
+ b \zeta_A e_A + \frac{1}{2} b \underline{\nu} L \\ 
& & +  \frac{1}{2} V^A \chi_{AB} e_B - \frac{1}{2} V^A \epsilon_A L  \\ 
& &  - a \zeta_A e_A +  \frac{1}{2} a \nu \Lu  \\ 
& & + b \underline{\xi} e_A - \frac{1}{2} b \underline{\nu} \Lu \\ 
& & + \frac{1}{2} V^A \underline{\chi}_{AB} e_B + \frac{1}{2} V^A \epsilon_A \Lu  \ . 
\eeas
Some of these expressions are zero. Above we show that $\nabla_L \Lu$ does not have any component along $\Lu$. In a similar manner it follows that 
$\nabla_{\Lu}L$ does not have any component along $L$. Straightforward computations along the lines as we do them above yield 
\bea
\nabla_k T & = &  \label{nablakT**1}  \\ 
& & 
b \zeta_A e_A +  \frac{1}{2} V^A \chi_{AB} e_B - \frac{1}{2} V^A \epsilon_A L 
  - a \zeta_A e_A 
 + \frac{1}{2} V^A \underline{\chi}_{AB} e_B + \frac{1}{2} V^A \epsilon_A \Lu  \ . \nonumber
\eea
All the connection coefficients except $\underline{\hat{\chi}}$ are $O(r^{-2})$, only $\underline{\hat{\chi}} = O(r^{-1})$, moreover the highest order terms of the traces of 
$\chi$ and $\underline{\chi}$ cancel. We take this into account as well as the decays for the vectorfields established above. 
Then we compute in the order of appearance for the terms on the right hand side of the previous equation 
\beas
& & (\nabla_k T) \cdot (aL + b \Lu + V)   =  \\ 
&  & 
b O(r^{-4 -\epsilon})  + O(r^{-4 -\epsilon}) + b O(r^{-4 -\epsilon}) + a O(r^{-4 -\epsilon}) +  O(r^{-3 -\epsilon}) + a O(r^{-4 -\epsilon})  \ . 
\eeas

Then we obtain 
\[
(\nabla_k T) \cdot (aL + b \Lu + V) = O(r^{-3 -\epsilon}) + a O(r^{-4 -\epsilon}) + b O(r^{-4 -\epsilon}) \ . 
\]

We derive 
\be
a+ b = T_b k^b = c_1 +  O(r^{-2 -\epsilon}) + a O(r^{-3 -\epsilon}) + b O(r^{-3 -\epsilon}) \ . 
\ee
The lower order terms multiplied by $a$ respectively $b$ can be absorbed into $a$ respectively $b$. 

Next, we do the corresponding computations with the vectorfield $K$ given in (\ref{intK12}), thus 
$K = \frac{1}{2} (\underline{u}^2 L + u^2 \Lu)$. It is 
\be \label{nablakK*1}
\nabla_k (K_b k^b) = (\nabla_k K_b) k^b = (\nabla_k K) \cdot k = (\nabla_k K) \cdot (aL + b \Lu + V)  \ . 
\ee
First, we consider $\nabla_k K$. 
\[
\nabla_k K = \underbrace{\frac{1}{2} \underline{u}^2 \nabla_k L + \frac{1}{2} u^2 \nabla_k \Lu}_{=: A} 
+ \underbrace{\underline{u} (\nabla_k \underline{u}) L 
+ u (\nabla_k u) \Lu}_{=: B} \ . 
\]
We investigate $A$, then $A \cdot k$. Then, we investigate $B$, then $B \cdot k$. 
The terms in $A$ emerge from the results above for $T$, but multiplied with the corresponding weights $\underline{u}^2$ and $u^2$, respectively. 
With (\ref{nablakT**1}) we find 
\beas
A  =  \frac{1}{2} \underline{u}^2 \nabla_k L + \frac{1}{2} u^2 \nabla_k \Lu & = &  
 b \underline{u}^2 \zeta_A e_A +  \frac{1}{2} \underline{u}^2 V^A \chi_{AB} e_B - \frac{1}{2} \underline{u}^2 V^A \epsilon_A L  \\ 
& & 
  - a u^2 \zeta_A e_A 
 + \frac{1}{2} u^2 V^A \underline{\chi}_{AB} e_B + \frac{1}{2} u^2 V^A \epsilon_A \Lu   \ . 
\eeas
It then follows directly for $A \cdot k$ (in order of appearance of the terms on the right hand side in the previous formula): 
\[
A \cdot k = 
b O(r^{-2})  + O(r^{-1 - \epsilon}) + b O(r^{-2 - \epsilon}) + a O(u^2 r^{-4}) +  O(u^2 r^{-3 - \epsilon}) + a O(u^2 r^{-4 - \epsilon})  
\]
Thus it is 
\[
A \cdot k  = O(r^{-1 - \epsilon}) + O(u^2 r^{-3 - \epsilon}) + a O(u^2 r^{-4})  + b O(r^{-2}) \ . 
\]

Let us focus on $B$. We have 
\[
B = \underbrace{\underline{u} (\nabla_k \underline{u}) L}_{=: B_1}
+ \underbrace{u (\nabla_k u) \Lu}_{=: B_2}
\]
The main part of the first term writes 
\[
\nabla_k \underline{u} = a \nabla_L \underline{u} + b \nabla_{\Lu} \underline{u} + V^A \nabla_A \underline{u} 
\]
whereas the main part of the second term is 
\[
\nabla_k u = a \nabla_L u + b \nabla_{\Lu} u + V^A \nabla_A u \ . 
\]
Then it is 
\beas
B_1 & = & \underline{u} (\nabla_k \underline{u}) L = 
a \underline{u} (\nabla_L \underline{u}) L + b \underline{u} (\nabla_{\Lu} \underline{u}) L + \underline{u} V^A (\nabla_A \underline{u}) L \\ 
B_2 & = &  u (\nabla_k u) L = a u (\nabla_L u) \Lu + b u (\nabla_{\Lu} u) \Lu + u V^A (\nabla_A u) \Lu \ . 
\eeas
Straightforward computations yield 
\beas
B \cdot k & = & 
ab \underline{u}(\nabla_L \underline{u}) L \Lu + 
b^2 \underline{u} (\nabla_{\Lu} \underline{u}) L \Lu + b \underline{u} V^A (\nabla_A \underline{u}) L \Lu \\ 
& & + 
a^2 u (\nabla_L u) \Lu L + ab u (\nabla_{\Lu} u) \Lu L + a u V^A (\nabla_A u) \Lu L \\ 
& = & 
ab \underline{u} (\nabla_L \underline{u}) L \Lu + ab u (\nabla_{\Lu} u) \Lu L \\ 
& = & ab \underline{u} O(r^{-2}) + ab u O(r^{-2}) \ . 
\eeas
The above holds because 
$0 = \nabla_L u = \nabla_{\Lu} \underline{u} = \nabla_A \underline{u} = \nabla_{\Lu} u = \nabla_A u$ and 
$\nabla_L \underline{u}$ as well as $\nabla_{\Lu} u$ are constant. 
We recall from above that $ab = \frac{1}{4} V^a V_a = O(r^{-2 - \epsilon})$.  This gives 
\[
B \cdot k = O(r^{-3 - \epsilon}) + O(u r^{-4 - \epsilon}) \ . 
\]

Putting the pieces together we find 
\beas
(\nabla_k K) \cdot k & = & A \cdot k + B \cdot k  \nonumber \\ 
& = & 
O(r^{-1 - \epsilon}) + O(u^2 r^{-3 - \epsilon}) + a O(u^2 r^{-4})  + b O(r^{-2})  \ . 
\eeas
Further we compute 
\be
u^2 a + \underline{u}^2 b = k_a K^a = c_2 + O(r^{- \epsilon}) +  O(u^2 r^{-2 - \epsilon}) + a O(u^2 r^{-3})  + b O(r^{-1}) \ .
\ee
Again we absorb the lower order terms in $a$ respectively $b$ into $a$ respectively $b$. 

Combining the results for $T$ and $K$, 
\bea
T_a k^a & = & a + b = c_1 + O(r^{-2 - \epsilon})   \label{Tk1} \\ 
K_a k^a & = & u^2 a + \underline{u}^2 b = c_2 + O(r^{- \epsilon}) \ .  \label{Tk2}
\eea

We are free to choose $c_1 = 1$. Then it is with (\ref{Tk1}) 
\[
a = 1 - b + O(r^{-2 - \epsilon}) \ . 
\]
Inserting in (\ref{Tk2}) yields 
\beas
u^2 (1 - b + O(r^{-2 - \epsilon}) ) + \underline{u}^2 b & = & c_2 + O(r^{- \epsilon}) \\ 
b(\underline{u}^2 - u^2 ) & = & c_2 - u^2 - u^2 O(r^{-2 - \epsilon}) + O(r^{- \epsilon}) \\ 
b & = & \frac{c_2 - u^2 +  u^2 O(r^{-2 - \epsilon}) + O(r^{- \epsilon}) }{\underline{u}^2 - u^2} \\ 
b & = & \frac{c_2 - u^2 +  u^2 O(r^{-2 - \epsilon}) + O(r^{- \epsilon})}{4r(u+r)}
\eeas
The latter equation holds because of $\underline{u} = u + 2r$. 

Thus we find that $b = O(r^{-2})$.  
As $r \to \infty$ it follows that $b \to 0$ and $a \to 1$. 

Then it follows that $k$ tends to $L$ along $C_u$ for $t \to \infty$. This ends the proof of convergence. 

{\bf Remark:} In the above arguments, there could in principle be terms involving $a^2$ and $b^2$. However, it can be easily shown that they vanish. The remaining terms involving $ab$ are estimated by identity (\ref{4abVV}). The quantities with $a$ and $b$ emerge with factors of lower order in $r$ and therefore are absorbed into $a$ and $b$ in the computations.

We recall that performing our experiment, we are at null infinity of our spacetime and receive gravitational wave signals traveling from the source along outward null hypersurfaces. 

The above shows that while a short time after the gravitational wave burst, the neutrinos following null curves may still 
fly into various directions, after some time the amount escaping towards the $\underline{L}$- and $V$-directions decay and the 
neutrino flow for later times will approach the null geodesic generated by $L$.

\section{Spacetime Structure}

Decomposing the Einstein-null-fluid equations with respect to the two main foliations, we derive the following equations.

\subsection{Equations for $t$-Foliation}

In this chapter, we give the most important equations to be used later in the paper. 

Given the Laplace operator $\triangle$ in $H$, its radial decomposition reads 
\be \label{Laplaceradial1}
\triangle  =  \nabla_N^2  + \slap +  tr \theta \nabla_N  + a^{-1} \nlap a \cdot \nlap \ . 
\ee

The second fundamental form $k$ obeys the equations 
\bea
tr k & = & 0   \label{k1}  \\
(curl \ k)_{ij} & = & H(W)_{ij} + \frac{1}{2} \epsilon_{ij}^{\ \ l} R_{0l}  \label{curlk1}  \\ 
(div \ k )_i & = & R_{0i}  \label{divk1}  \ . 
\eea
We have to take into account that the Ricci curvature $\bar{R}_{ij}$ in the spacelike hypersurfaces $H_t$ is composed as in the following formula, with $R_{\alpha \beta}$ denoting the spacetime Ricci curvature of $M$. 
\be \label{RicciHt1}
\bar{R}_{ij} \ = \ k_{ia} k^{a}_{j} + E(W)_{ij} + \frac{1}{2} g_{ij} R_{00} + \frac{1}{2} R_{ij} \ . 
\ee

In particular, the components $\delta, \epsilon, \eta$ of $k$ satisfy: 
\bea
\dlap \epsilon \ & = & \ 
- \nabla_N \delta - \frac{3}{2} tr \theta \delta + \hat {\eta} \cdot \hat{\theta} - 2 (a^{-1} \nlap a ) \cdot \epsilon 
\nonumber \\ 
\ & & \ 
-  2 \pi   T_{\underline{L} \underline{L}} + 2 \pi 
T_{LL}  \\ 
\clap \   \ \epsilon  \ & = & \ 
\sigma (W) + \hat{\theta} \wedge \hat{\eta}  \\ 
\nlap_N \epsilon + tr \theta \epsilon \ & = & \ 
- \nlap \delta - \hat{\theta} \cdot \epsilon + \frac{3}{2} (a^{-1} \nlap a) \cdot \delta - \hat{\eta} \cdot  (a^{-1} \nlap a) \nonumber  \\ 
\ & & \ 
+ \frac{1}{2} (\beta - \underline{\beta}) 
+ 2\pi (T_{A \Lu} - T_{A L})   \\ 
\dlap \hat{\eta} \ & = & \ 
- \frac{1}{2} \nlap \delta + \hat{\theta} \cdot \epsilon - \frac{1}{2} tr \theta \cdot \epsilon \nonumber \\ 
\ & & \ 
+ \frac{1}{2} (\beta - \underline{\beta}) 
+ 8 \pi ( T_{A \Lu} - T_{A L}  ) \\ 
\nlap_N \hat{\eta} + \frac{1}{2} tr \theta \hat{\eta} \ & = & \ 
\frac{3}{2} \delta \cdot \hat{\theta} + \frac{1}{2} \nlap \hat{\otimes} \epsilon + (a^{-1} \nlap a)  \hat{\otimes} \epsilon  \nonumber \\ 
\ & & \ 
+ \frac{1}{4} (\alpha - \underline{\alpha}) \ \ . 
\eea

\subsection{Null Structure Equations}

The main quantities to derive nonlinear memory are the shears $\hat{\chi}$ and $\hat{\underline{\chi}}$. Equations for the latter on $2$-surfaces are coupled to evolution equations of the corresponding traces. 
Propagation equations of $tr \underline{\chi}$ and $tr \chi$ with respect to $l$-pair: 
\begin{eqnarray}
\frac{dtr\underline{\chi }}{ds}  &=& - \frac{1}{2}tr\chi tr\underline{\chi }  - 2
\underline{\mu }+2\left \vert \zeta \right \vert ^{2} \label{nullstruct1} \\
\frac{dtr\chi }{ds} &=& - \frac{1}{2}\left( tr\chi \right) ^{2}   -\left \vert 
\widehat{\chi }\right \vert ^{2} -  
8 \pi T_{LL} .   \label{nullstruct2}
\end{eqnarray}

The Gauss equation reads 
\begin{equation*}
K=-\frac{1}{4}tr\chi tr\underline{\chi }+\frac{1}{2}\widehat{\chi }\cdot 
\underline{\widehat{\chi }}-\rho \left( W\right) - 4 \pi T_{L \Lu}. 
\end{equation*}%

Define the function $\underline{\mu}$ as 
\[
\underline{\mu} = - \dlap \underline{\zeta} + \frac{1}{2} \hat{\chi} \cdot \underline{\hat{\chi}} - \rho (W) - 4 \pi T_{L \Lu} . 
\]

The latter, with the help of the Gauss curvature $K$, can be written as 
\begin{equation}
\underline{\mu }=- \dlap \underline{\zeta }+K+\frac{1}{4}tr\chi tr%
\underline{\chi }.  \label{Gaussmassaspect}
\end{equation}%

The null Codazzi and conjugate null Codazzi equations read 
\bea
\dlap \hat{\chi} \ & = & \  - \hat{\chi} \cdot \zeta + \frac{1}{2} (\nlap tr \chi + \zeta tr \chi) - \beta -  8 \pi T_{A L} \\ 
\dlap \hat{\underline{\chi}} \ & = & \ \hat{\underline{\chi}} \cdot \zeta + \frac{1}{2} (\nlap tr \underline{\chi} - \zeta tr \underline{\chi}) + \underline{\beta} + 
8 \pi T_{A \Lu} \ \ . 
\eea

Useful identities 
\bea 
\frac{dr}{dt} & = &  \frac{1}{2} r \ \overline{a \phi tr \chi}   \label{drdt1} \\ 
\frac{dr}{du} & = &  \frac{1}{2} r \ \overline{a  tr \theta}   \label{drdu1}  . 
\eea

\section{Null Infinity} 

\subsection{Null Asymptotic Limits}

\begin{The} \label{conclcurvandfieldcompts1}
Let $C_{u}$ denote any null hypersurface. 
Then 
the normalized curvature components $r\underline{\alpha }\left( W\right) $, $r^{2}\underline{\beta }\left(
W\right) $, $r^{3}\rho \left( W\right) $, $r^{3}\sigma \left( W\right) $, 
and normalized energy-momentum components 
$r^2 T_{\Lu \Lu}, r^4 T_{L \Lu}, r^4 T_{AB}  , r^3 T_{A \Lu}, r^5 T_{A L}$ 
have limits as $t\rightarrow \infty $. That is 
\begin{eqnarray*}
\lim_{C_{u},t\rightarrow \infty }r\underline{\alpha }\left( W\right)
&=&A_{W}\left( u,\cdot \right) ,\, \ \ \ \ \ \ \ \ \ \ \ \
\lim_{C_{u},t\rightarrow \infty }\,r^{2}\underline{\beta }\left( W\right)
=B_{W}\left( u,\cdot \right) \\
\lim_{C_{u},t\rightarrow \infty }r^{3}\rho \left( W\right) &=&P_{W}\left(
u,\cdot \right) ,\, \ \ \ \ \ \ \ \ \ \ \ \ \lim_{C_{u},t\rightarrow \infty
}r^{3}\sigma \left( W\right) =Q_{W}\left( u,\cdot \right) \\
\lim_{C_{u},t\rightarrow \infty }r^2 T_{\Lu \Lu} 
&=& T_{\Lu \Lu}^* \left( u,\cdot \right) , \\
\lim_{C_{u},t\rightarrow \infty }r^3 T_{A \Lu} 
&=& T_{A \Lu}^* \left( u,\cdot \right) , \\
\lim_{C_{u},t\rightarrow \infty }r^4 T_{L \Lu} 
&=& T_{L \Lu}^* \left( u,\cdot \right) , \\
\lim_{C_{u},t\rightarrow \infty }r^{4} T_{AB} &= & T_{AB}^*  \left(
u,\cdot \right) ,\  \\ 
\lim_{C_{u},t\rightarrow \infty }r^{5} T_{A L} &= & T_{A L}^*  \left(
u,\cdot \right) ,\ 
\end{eqnarray*}
where the limits are on $S^{2}$ and depend on $u$. Moreover, these limits satisfy 
\begin{eqnarray*}
\left| A_{W}\left( u,\cdot \right) \right| &\leq &C\left( 1+\left| u\right|
\right) ^{-5/2}\, \, \ \ \ \ \ \ \ \ \ \ \ \ \left| B_{W}\left( u,\cdot
\right) \right| \leq C\left( 1+\left| u\right| \right) ^{-3/2} \\
\left| P_{W}\left( u,\cdot \right) -\overline{P}_{W}\left( u\right) \right|
&\leq &\left( 1+\left| u\right| \right) ^{-1/2}\, \, \quad \quad \quad \quad
\, \left| Q_{W}\left( u,\cdot \right) -\overline{Q}_{W}\left( u\right)
\right| \leq \left( 1+\left| u\right| \right) ^{-1/2} \\
T_{\Lu \Lu }^* \left( u,\cdot \right)  &\leq &C\left( 1+\left| u\right|
\right) ^{-4} \\ 
T_{A \Lu}^* \left( u,\cdot \right)  &\leq &C\left( 1+\left| u\right|
\right) ^{-3} \\ 
T_{L \Lu}^* \left( u,\cdot \right) &\leq &\left( 1+\left| u\right|
\right) ^{-2} \\ 
T_{AB}^* \left( u,\cdot \right) &\leq &\left( 1+\left| u\right|
\right) ^{-2} \\ 
T_{A L}^* \left( u,\cdot \right) &\leq &\left( 1+\left| u\right|
\right) ^{-1}   \, \, \ \
\end{eqnarray*}
and 
\begin{equation*}
\lim_{u\rightarrow -\infty }\overline{P}_{W}\left( u\right) =0,\, \ \ \ \ \
\ \lim_{u\rightarrow -\infty }\overline{Q}_{W}\left( u\right) =0.
\end{equation*}
\end{The}
{\bf Proof:} 
Whereas the proof of the properties of the Weyl tensor components is along the lines of \cite{sta}, we 
establish the results for the null fluid above. The estimates of this theorem follow directly.

The following theorem shows behavior of the shears and the fundamental relation between the shears and the curvature, which is in accordance with the picture found by Christodoulou-Klainerman in \cite{sta} and by Zipser in \cite{zip}, \cite{zip2}. In our setting here, we prove that the null fluid terms do not change these equations. 

\begin{The} \label{conclSigma1}
Consider the null hypersurface $C_{u}$. 
The normalized shear $r^{2} \widehat \chi ^{\prime }$
tends to the following limit as $t\rightarrow \infty $: 
\begin{equation*}
\Sigma \left( u,\cdot
\right)
= 
\lim_{C_{u},t\rightarrow \infty }r^{2}  \widehat \chi ^{\prime } \ . 
\end{equation*}
The limit $\Sigma $ is a symmetric traceless covariant 2-tensor on $S^{2}$ that 
depends on $u$.
\end{The}

The proof is the same as in \cite{sta} because the 
propagation equation is not affected by the extra terms from the energy-momentum tensor of the null fluid. 
\begin{equation*}
\frac{d\widehat{\chi }_{AB}}{ds}=-tr\chi \widehat{\chi }_{AB}-\alpha
(W)_{AB}.
\end{equation*}

\begin{The} \label{conclXi1}
Consider any null hypersurface $C_{u}$. 
The limit of $r\widehat{\eta }$ exists as $%
t\rightarrow \infty $, in particular 
\begin{equation*}
\Xi \left( u,\cdot \right) = 
\lim_{C_{u},t\rightarrow \infty }r\widehat{\eta } \ . 
\end{equation*}
The limit $\Xi $ is a symmetric traceless 2-covariant tensor on $S^{2}$ that 
depends on $u$ and obeys 
\begin{equation*}
\left| \Xi \left( u,\cdot \right) \right| _{\overset{\circ }{\gamma }}\leq
C\left( 1+\left| u\right| \right) ^{-3/2}.
\end{equation*}
In addition, the following holds: 
\begin{equation*}
\lim_{C_{u},t\rightarrow \infty }r\widehat{\theta }=-\frac{1}{2}%
\lim_{C_{u},t\rightarrow \infty }r\widehat{\underline{\chi }}^{\prime }=\Xi
\end{equation*}
and
\begin{eqnarray}
\frac{\partial \Xi }{\partial u} &=&-\frac{1}{4}A_{W}.  \label{Xiu*1} \\ 
\frac{\partial \Sigma }{\partial u} &=&-\Xi   \label{Sigmau*1}  
\end{eqnarray}
\end{The}

To prove this, we take into account the decay behavior of the energy-momentum tensor components for the null fluid. The argument 
is along the lines of the proof of conclusion 17.0.3 in \cite{sta}.

In the proof of theorem \ref{displ*1} below we use a fact on the limit of $tr \chi'$ which we want to establish now. 
For that purpose we define 
\[
H = \lim_{C_u, t \to \infty} \Big( r^2 (tr \chi' - \frac{2}{r}) \Big)
\]

\begin{Lem} \label{lemH1}
The following holds for the function $H$: 
\begin{equation} \label{H1} 
\frac{\partial H}{\partial u} =0 
\end{equation}
\begin{equation}
\bar H = 0  \label{H2}
\end{equation}
\end{Lem}
{\bf Proof:} 
At the beginning, we want to remind the reader that the equivalent statement for the EV equations is shown in conclusion 17.0.5 and in lemma 17.0.1 of \cite{sta}. 
There the authors use 
\[ \nabla_N tr \chi' +\frac{1}{2}  \chi'  = O(r^{-3}). \]

In our new setting for the Einstein null fluid, it follows in a straightforward manner that the additional terms due to the null fluid 
are of order $O(r^{-3})$.
Then by the same argument as in \cite{sta}, above equation (\ref{H1}) follows in the presence of a null fluid.

In order to prove equation (\ref{H2}) in the null fluid case, we recall the EV situation from lemma 17.0.1 in \cite{sta}. 
One has to show that 
$ r^2 \bar \delta  $ converges to $2 M(u)$. 
Now, Proposition 4.4.4 in \cite{sta} says that

\[  4 \pi r^3 \bar \delta  = \int_{u_0}^u du'( \int_{S_{t,u}}  ar \hat \theta \cdot \hat \eta  - \frac{1}{2} \kappa(\delta - \bar \delta) - r a^{-1}  \nlap a \cdot \epsilon + r (div k)_N ) \]

Along the lines of proof of lemma 17.0.1 in \cite{sta}, it follows 
\[   \int_{S_{t,u}}  ar \hat \theta \cdot \hat \eta  - \frac{1}{2} \kappa(\delta - \bar \delta) - r a^{-1}  \nlap a \cdot \epsilon  = r \int_{S^2} |\Xi|^2 d\mu _{\overset{\circ }{\gamma }} +O(1)\]

The constraints of the Einstein null fluid equations give 
\begin{equation}  \label{dvikn} (div k)_N =  R_{0N}  =  8 \pi T_{0N} = 2 \pi (T_{\Lu \Lu} - T_{LL})  . \end{equation}
From equation (\ref{dvikn}), we deduce 
\[   \int_{S_{t,u}}  r (div k)_N = 2 \pi r \int_{S^2} T_{\Lu \Lu} d\mu _{\overset{\circ }{\gamma }} +O(1) \ .\]
As a consequence, we infer that 
\[ r \bar \delta  = \frac{2}{r^2} \int_{u_0}^u  r \frac{\partial }{\partial u}  m(t,u) +  O(r^{-1})\]
This concludes the main part where the null fluid components enter. 
The remaining steps follow easily.

\subsection{Bondi Mass}

Next, we study the Bondi mass in our setting. 
First, we introduce the Hawking mass $m$ inclosed by a $2$-surface $S_{t,u}$ as in \cite{chrmemory} to be 
\be  \label{hawking mass}
m(t,u) = \frac{r}{2} \big( 
1 + \frac{1}{16 \pi} \int_{S_{t,u}} tr \chi tr \underline{\chi}  
 \big)  . 
\ee

We first investigate $\frac{\partial}{\partial t} m(t,u)$ and then $\frac{\partial}{\partial u} m(t,u)$. From the limiting behavior of the former we conclude a convergence result of $m(t,u)$ to the Bondi mass, and from the limiting behavior of the latter we compute the Bondi mass loss formula.

Consider the null structure equations ((\ref{nullstruct1}), (\ref{nullstruct2})). 
To compute $\frac{d}{ds} (tr \chi tr \underline{\chi})$ we add $tr \underline{\chi} \cdot$(\ref{nullstruct2}) and $tr \chi \cdot$(\ref{nullstruct1}), 
which yields 
\begin{eqnarray*}
\frac{d}{ds}( tr\chi tr\underline{\chi } ) &=&-  \left( tr\chi \right)^2 tr\underline{
\chi } -tr\underline{\chi }\left \vert \widehat{\chi }\right \vert ^{2} -2\underline{\mu }tr\chi +2tr\chi \left \vert \zeta \right
\vert ^{2} - 8 \pi tr 
\underline{\chi } T_{LL}  . 
\end{eqnarray*}%
Then we derive 
\begin{eqnarray}
\frac{\partial }{\partial t}\int_{S_{t,u}}tr\chi tr\underline{\chi }
&=& \int_{S_{t,u}}a\phi \left( -tr\underline{\chi } |  \widehat{\chi } |^2 + 2tr\chi \left \vert \zeta \right \vert^{2} 
- 8 \pi tr \underline{\chi} T_{LL}  \right)  \label{t-integrate trxtrxbar} \\ 
& & -2\int_{S_{t,u}}a\phi \underline{\mu }tr\chi .   \notag
\end{eqnarray}

Next, we use (\ref{Gaussmassaspect}) to integrate $\underline{\mu}$ on $S_{t,u}$. Applying Gauss-Bonnet yields 
\be \label{mmu1}
\int_{S_{t,u}} \underline{\mu} = \int_{S_{t,u}} 4 \pi 
\Big( 
1 + \frac{1}{16 \pi} \int_{S_{t,u}} tr \chi tr \underline{\chi} 
  \Big) 
  = \frac{8 \pi}{r} m . 
\ee

Finally, from (\ref{t-integrate trxtrxbar}) and (\ref{mmu1}), using identity (\ref{drdt1}), 
we conclude 
\bea
\frac{\partial}{\partial t} m(t,u) = 
\frac{r}{8 \pi} \int_{S_{t,u}} a \phi \Big( 
- \frac{1}{4} tr\underline{\chi } |  \widehat{\chi } |^2 + \frac{1}{2} tr\chi \left \vert \zeta \right \vert^{2} 
- 2 \pi tr \underline{\chi} T_{LL} 
\Big)  \nonumber  \\ 
- \frac{r}{16 \pi} \int_{S_{t,u}} ( a \phi tr \chi - \overline{a \phi tr \chi}  ) \underline{\mu} . \label{dmdt2}
\eea
Let us have a look at the terms on the right hand side. 
From the fact that $\underline{\mu} = O(r^{-3})$ it follows that the integrand on the second line of (\ref{dmdt2}) is $O(r^{-5})$. 
Moreover, all the integrands on the first line of (\ref{dmdt2}) are also $O(r^{-5})$ or higher order. 
This leads to the conclusion 
\[
\frac{\partial}{\partial t} m(t,u) = O(r^{-2}) .
\]
Then we keep $u$ fixed and observe $m(t,u)$ to reach its limit $M(u)$ as $t \to \infty$. This limit $M(u)$ is called the Bondi mass, and it is defined for 
each null hypersurface $C_u$. 
Thus, in each $C_u$ as $t \to \infty$ the Hawking mass $m(t,u)$ equals the Bondi mass $M(u)$ plus terms decaying like $O(r^{-1})$. 
Note that the null fluid term on the right hand side of (\ref{dmdt2}) decays fast enough not to interfere with the `purely geometrical' parts. 
Thus, we have proven the following theorem. 
\begin{The}
On any null hypersurface $C_u$ the Hawking mass $m(t,u)$ tends to the Bondi mass $M(u)$ as $t \to \infty$, in particular it is:  
$m(t,u) = M(u) + O(r^{-1})$.  
\end{The}

Having understood how the Hawking mass tends to the Bondi mass, the next question is how the mass changes going from one null hypersurface to another. 
From above we have $m = \frac{r}{8 \pi} \int_{S_{t,u}} \underline{\mu}$. 
Our goal is to derive the Bondi mass loss formula. 
In order to do so, we turn to $\frac{\partial}{\partial u} m(t,u)$ and write 
\beas
\frac{\partial}{\partial u} m(t,u) & = & \underbrace{\frac{1}{16 \pi} \Big( \int_{S_{t,u}} \underline{\mu} \Big) \cdot r}_{= \frac{1}{2}m} \cdot \overline{a tr \theta} + 
\frac{r}{8 \pi} \underbrace{\frac{\partial}{\partial u} \int_{S_{t,u}} \underline{\mu}}_{= \int_{S_{t,u}} a (\nabla_N \underline{\mu} + tr \theta \underline{\mu} ) } \\ 
& = & 
\frac{1}{2} m \cdot    \overline{a tr \theta} +  \frac{r}{8 \pi}  \int_{S_{t,u}} a (\nabla_N \underline{\mu} + tr \theta \underline{\mu} ) . 
\eeas
The last integrand can be written as follows using $e_4 = a^{-1}(T+N)$ and $e_3 = a(T - N)$: 
\[
 a (\nabla_N \underline{\mu} + tr \theta \underline{\mu} ) = \frac{1}{2} a^2 (D_4 \underline{\mu} + tr \chi \underline{\mu} )
 - \frac{1}{2} (D_3 \underline{\mu} + tr \underline{\chi} \underline{\mu}  ) . 
\]
We compute 
\beas
D_4 \underline{\mu} + tr \chi \underline{\mu} & = &  O(r^{-4}) \\ 
D_3 \underline{\mu}  + tr \underline{\chi} \underline{\mu} & = & - \frac{1}{4} tr \chi | \underline{\hat{\chi}} |^2 
- 4 \pi tr \chi T_{\Lu \Lu} + O(r^{-4}) . 
\eeas
We derive 
\[
\frac{\partial}{\partial u} m(t,u) = 
\frac{r}{64 \pi} \int_{S_{t,u}} tr \chi \Big(  | \underline{\hat{\chi}} |^2 + 16 \pi  T_{\Lu \Lu}  \Big) + O(r^{-4}) . 
\]
In order to derive the limit yielding the Bondi mass-loss formula, we need to check the limits of each term in the expression $\frac{\partial m(t,u)}{\partial u}$ for the Hawking mass. 
First, we recall that for each $u$, $\phi^*_{t,u}$ denotes a diffeomorphism from the unit sphere $S^2$ to $S_{t,u}$. 
Then by arguments along the lines as in \cite{sta} it follows that for each $u$ 
as $t \to \infty$ the metric $\tilde{\gamma} = \phi^*_{t,u} (r^{-2} \gamma)$ converges to the standard metric $\overset{\circ }{\gamma }$ on $S^2$. 
It follows in a straightforward manner that for each $u$ as $t \to \infty$, $r tr \chi$ converges to $2$, and $r \underline{\hat{\chi}}$ converges to $-2 \Xi$. 
Recall that $T_{\Lu \Lu}^*$ is positive. 
Taking the limits we obtain the Bondi mass-loss formula: - See \cite{bondibm} for the first appearance of Bondi mass-loss. 
\[
\frac{\partial}{\partial u} M(u) = \frac{1}{8 \pi} \int_{S^2}  \Big(  | \Xi |^2 + 4 \pi   T_{\Lu \Lu}^*  \Big) 
d \mu_{\overset{\circ }{\gamma }} . 
\]
This term is positive and integrable in $u$, from which it follows that the Bondi mass $M(u)$ is a non-decreasing function of $u$. Moreover, it has finite 
limits $M(- \infty)$ and $M(\infty)$ as $u$ tends to $- \infty$ and $+ \infty$ respectively. 
From (\ref{mmu1}) it follows that $M(- \infty)=0$ and that $M(\infty)$ is the total mass. 
We have therefore proven the next theorem: 
\begin{The} \label{Bondimassloss*1}
The Bondi mass $M(u)$ obeys the following Bondi mass-loss formula: 
\[
\frac{\partial }{\partial u} M(u) = \frac{1}{8 \pi} \int_{S^2}  \Big(  | \Xi |^2 + 4 \pi  T_{\Lu \Lu}^* \Big) 
d \mu_{\overset{\circ }{\gamma }} 
\]
where $d \mu_{\overset{\circ }{\gamma }}$ denotes the area element of the standard unit sphere $S^2$. 
\end{The}
We compare this result with the corresponding formulas in the purely gravitational case and in the electromagnetic case. 
See \cite{sta}, \cite{chrmemory}, \cite{zip2}, \cite{1lpst1}. Thus, we find that the energy-momentum tensor of the null fluid describing neutrino radiation contributes to the change of the Bondi mass through 
the term $\frac{1}{2} \int_{S^2}  T_{\Lu \Lu}^* d \mu_{\overset{\circ }{\gamma }}$. 

The behavior of $\Xi$ and $T_{\Lu \Lu}^*$ in $u$ are consequences of theorems \ref{conclcurvandfieldcompts1} and \ref{conclXi1}. 
Now, we define the function 
\be
F = \frac{1}{8} \int_{- \infty}^{+ \infty} \Big( | \Xi |^2 + 4 \pi  T_{\Lu \Lu}^*   \Big) du .  
\ee
We then find the total energy radiated to infinity in a given direction per unit solid angle to be $\frac{F}{4 \pi}$. We note that neutrino radiation contributes through its corresponding null component limit $T_{\Lu \Lu}^*$.

\subsection{Permanent Displacement Formula}

The difference $\Sigma^+ - \Sigma^-$ governs 
the permanent displacement formula of test particles in a gravitational wave detector. 
In this section, we prove a theorem for $\Sigma^+ - \Sigma^-$ in the case of neutrino radiation described by a null fluid in the Einstein equations. 

The theorem we prove in this chapter employs the full and rich geometric-analytic structure of our spacetime. The emerging result 
is then related to experiment in the last part of the present article. 

At this point, we emphasize that $T_{\Lu \Lu}^*$ is positive. 

\begin{The} \label{displ*1}
Denote by $\Sigma^+ (\cdot)$ the limit $\Sigma^+ (\cdot) = \lim_{u \to \infty} \Sigma (u, \cdot)$ and by 
$\Sigma^- (\cdot)$ the limit $\Sigma^- (\cdot) = \lim_{u \to - \infty} \Sigma (u, \cdot)$. 
Let 
\be \label{Thm*FXiN*1}
F (\cdot)  =   \int_{- \infty}^{\infty} 
\big( 
\mid \Xi (u, \cdot) \mid^2 + 4 \pi T_{\Lu \Lu}^* (u, \cdot) 
\big)
du  \ \ . 
\ee
Also, let 
$\Phi$ be the solution with $\bar{\Phi} = 0$ on $S^2$ of the equation 
\[
\stackrel{\circ}{\slap} \Phi = F - \bar{F}   \ \ . 
\]
Then 
$\Sigma^+ - \Sigma^-$ is determined by the following equation on $S^2$. 
\be \label{Thm*divSigma+-*2}
\stackrel{\circ}{\dlap} (\Sigma^+ - \Sigma^-) = \stackrel{\circ}{\nlap} \Phi \ \ . 
\ee
\end{The}

{\bf Proof:}  
First, one has to check on the limits of $\Sigma$. 
Theorem \ref{conclXi1}, 
equation (\ref{Sigmau*1}) ensures that 
$\Sigma$ tends to limits $\Sigma^+$ as $u \to \infty$ and $\Sigma^-$ as $u \to -
\infty$. Moreover, one has 
\[
\Sigma (u) - \Sigma^-  =  - \int_{- \infty}^u \Xi (u') du' 
\]

as well as 
\[
\Sigma^+ - \Sigma^- = - \int_{- \infty}^{\infty} \Xi (u') du'  \ \ . 
\]

Let us now explore how to get the limiting equation at null infinity for $\Sigma$. 
For this purpose, we focus on the normalized null Codazzi equation 
\be \label{*normalizednullcodazzi**1}
(\dlap \hat{\chi})_A  - \frac{1}{2} \nlap_A tr \chi + \epsilon_B \hat{\chi}_{AB} -
\frac{1}{2} \epsilon_A tr \chi  =  
- \beta (W)_A - 8 \pi T_{A L} \ . 
\ee
Then we multiply equation (\ref{*normalizednullcodazzi**1}) by $r^3$ and take the limit as 
$t \to \infty$ on $C_u$. 
We also introduce 
\[ E= \lim_{C_{u},t\rightarrow \infty }\left( r^2 \epsilon \right ) \ . \]
We derive thereby the limiting equation on $S^2$: 
\be \label{*circSigma**2}
 \stackrel{\circ}{\dlap} \Sigma = \stackrel{\circ}{\nlap} H+E \ \ , 
\ee 

This structure shows the same as in the EV case, which is proven in \cite{sta} p. 510, conclusion 17.0.8. 

Next, from our result on $H$ in lemma \ref{lemH1} equation (\ref{H1}) we obtain 
\be \label{*circSigma**3}
\stackrel{\circ}{\dlap} (\Sigma) = E \ \ . 
\ee

Thus, the next task is to investigate $E$ at null infinity through its limiting Hodge system on $S^2$. We therefore study the Hodge system for $\epsilon$: 
\bea
\dlap \epsilon \ & = & \ 
- \nabla_N \delta - \frac{3}{2} tr \theta \delta + \hat {\eta} \cdot \hat{\theta} 
\nonumber \\ 
\ & & \ 
- 2 (a^{-1} \nlap a ) \cdot \epsilon -  2 \pi   T_{\Lu \Lu}  + 
 2 \pi 
 T_{LL}   \label{systeps1} \\ 
\clap \   \ \epsilon  \ & = & \ 
\sigma (W) + \hat{\theta} \wedge \hat{\eta} \ \ .  \label{systeps2}
\eea
To derive equation (\ref{systeps1}), we consider (\ref{divk1}) and write for the normal component. 
\bea
R_{0N} = (div k)_N & = & \nabla_N k_{NN} + \gamma^{AB} \nabla_B k_{NA}  \nonumber \\ 
& = & 
\nabla_N \delta + 2 (a^{-1} \nlap a) \cdot \epsilon + \dlap \epsilon + \frac{3}{2} \delta \cdot tr \theta - \hat{\eta} \cdot \hat{\theta}  \label{divkN1} \ . 
\eea
From here with (\ref{Einsteinnullfluid1}) we compute directly and obtain (\ref{systeps1}). 
Further, we use (\ref{curlk1}) to find the $\clap \ $ equation (\ref{systeps2}). The latter in fact 
coincides with the one obtained by
Christodoulou and Klainerman in \cite{sta}, whereas 
the $\dlap $ equation (\ref{systeps1}) contains the extra terms $ T_{LL}  $ and $ T_{\Lu \Lu}$ from 
the null fluid.

In the $\dlap / \clap \ $ system ((\ref{systeps1}), (\ref{systeps2})) we make use of underlying structures when taking the limit 
on $C_u$ as $t \to \infty$. 
To extract these structures, we introduce 
$\Psi$, $\Psi'$ as follows:  
\bea
\triangle \Psi & = & r \mid \hat{\eta} \mid^2  -  2 \pi r T_{\Lu \Lu}  \label{trianglePsi1}  \\ 
\triangle \Psi' & = & - r a^{-1} \lambda \big( \mid \hat{\eta} \mid^2 -
\overline{\mid \hat{\eta} \mid^2} \big) 
 +  2 \pi r^2 a^{-1} \big( a  \Dlap_4  T_{\Lu \Lu}    -
\overline{a  \Dlap_4   T_{\Lu \Lu}   }  \big) 
\label{trianglePsip1}
\eea
The reader may want to compare this with the formulas in the EV case by 
Christodoulou and Klainerman in \cite{sta}, chapter 11.2,
(11.2.2b) and (11.2.7b)
which read 
\bea
\triangle \Psi & = & r \mid \hat{\eta} \mid^2  \label{trianglePsi2} \\ 
\triangle \Psi' & = & - r a^{-1} \lambda \big( \mid \hat{\eta} \mid^2 -
\overline{\mid \hat{\eta} \mid^2} \big)  \label{trianglePsip2}  \ \ . 
\eea
In our new situation of neutrino radiation given by the null fluid, we compute the limits as 
\bea 
\lim_{C_u, t \to \infty} \Psi = \mathbf{\Psi} \ \ \ \ \quad & & \ \ \ \ \quad 
\lim_{C_u, t \to \infty} \Psi' = \mathbf{\Psi'}  \nonumber \\ 
\lim_{C_u, t \to \infty} r \nabla_N \Psi = \Omega(u, \cdot )  \ \ & & \ \ 
\lim_{C_u, t \to \infty} r \nabla_N \Psi' = \Omega'(u, \cdot ) 
\label{limitsPsiOmega*1}.
\eea 

We proceed by investigating $\nabla_N \delta$ in equation (\ref{systeps1}). 
Writing 
the following equation for $\nabla_N \delta$ and comparing it to \cite{sta}, chapter 17, (17.0.12c), we note that 
our formula (\ref{nablaNdelta*1}) differs from that by the extra term from the null fluid. 
\bea
\nabla_N \delta - \hat{\theta} \cdot \hat{\eta} +  2 \pi T_{\Lu \Lu}   \ & = & \ 
- 2r^{-3} (\nabla_N r) p + r^{-2} \nabla_N p - r^{-2} (\nabla_N r) \nabla_N \Psi +
r^{-1} \nabla_N^2 \Psi  \nonumber \\ 
\ & = & \ 
- \hat{\chi} \cdot \hat{\eta} - r^{-1} \slap \Psi - r^{-2} 
\big( 
r tr \theta + a^{-1} \lambda \big) \nabla_N \Psi   \nonumber  \\ 
\ & & \ 
- r^{-1} a^{-1} \nlap a \cdot \nlap \Psi + r^{-2} \nabla_N p - 2 r^{-3} a^{-1}
\lambda p  \label{nablaNdelta*1}
\eea
with 
\[
p = r \nabla_N q + q' + \Psi' \ \ \ \mbox{ and } \ \ \ 
p = r (r \delta - \nabla_N \Psi) \ . 
\]
In a straightforward manner, along the lines of the argument in \cite{sta} and also used in \cite{zip2}, we show that 
\be \label{triangleq*1}
\triangle q \ = \ r (\mu - \overline{\mu}) +I 
\ee
with 
\beas
I \ & = & \ \frac{1}{2} \ ^{(rN)} \hat{\pi}_{ij} k_{ij} - \triangle \Psi -  2 \pi r  T_{\Lu \Lu } +  2 \pi r  T_{LL}    \\ 
\ & = & \ 
r \hat{\chi} \cdot \hat{\eta} - \kappa \delta - 2 r a^{-1} \nlap a \cdot \epsilon +
 2 \pi r T_{LL} \  
\eeas
Recall the mass aspect function $\mu$ to be 
\[  \mu=   - \widehat \chi   \cdot \hat \eta   - \rho(W) \ .\]

In the next step, we make use of the radial decomposition of $\triangle$ given in formula (\ref{Laplaceradial1}) above. 
Direct conclusions from the last equations yield 
\bea 
\triangle q \ & = & \ 
\nabla_N^2 q +  tr \theta \nabla_N q + \slap q + a^{-1} \nlap a \cdot \nlap q  
\nonumber \\ 
\ & = & \ 
-r(\rho - \bar{\rho}) - r \overline{\hat{\chi} \cdot \hat{\eta}} - \kappa \delta - 2
r a^{-1} \nlap a \cdot \epsilon 
+ 2 \pi r  T_{LL}  \label{triangle*2} 
\eea

Now, first we substitute for $\nabla_N p$ from (\ref{triangle*2}) in (\ref{nablaNdelta*1})
and then 
the resulting terms from (\ref{nablaNdelta*1}) in (\ref{systeps1}) to obtain 
\bea 
\dlap \epsilon \ & = & \ 
\rho - \bar{\rho} + \hat{\chi} \cdot \hat{\eta} - \overline{\hat{\chi} \cdot
\hat{\eta}}  \nonumber \\ 
& & 
+ r^{-1} \slap \Psi - r^{-2} \nabla_N \Psi' - r^{-3} a^{-1} \lambda \Psi' + l.o.t. 
\label{systeps3} \\ 
\clap \   \ \epsilon  \ & = & \ 
\sigma (W) + \hat{\theta} \wedge \hat{\eta}  \label{systeps4}
\eea

Then we multiply equations (\ref{systeps3}) and  (\ref{systeps4}) by $r^3$ and take the limits on $C_u$ as $t \to \infty$. 
We thereby derive the following limiting equations for $E$ on $S^2$. That is the Hodge system for $E$ at null infinity. 
\bea 
\stackrel{\circ}{\clap} \ \ E \ & = & \ Q + \Sigma \wedge \Xi  \label{systeps*5*1}  \\ 
\stackrel{\circ}{\dlap} E \ & = & \ P - \bar{P} + \Sigma \cdot \Xi - 
\overline{\Sigma \cdot \Xi}  \nonumber \\ 
\ & & \ 
+ \stackrel{\circ}{\slap} \Psi - \Psi' - \Omega' \ \ .  \label{systeps*5}
\eea 
We investigate 
the limits as $u \to + \infty$ and $u \to - \infty$. 
Taking into account the above equations for $\epsilon$ and $E$, applying theorems 
\ref{conclXi1} and 
\ref{conclcurvandfieldcompts1} 
it follows that 
$E$ tends to a limit $E^+$ as $u \to + \infty$ and to $E^-$ as $u \to
- \infty$. \\ \\ 
Similar arguments as in \cite{sta}, chapter 17, 
yield 
\[
\stackrel{\circ}{\clap} \ \ (E^+ - E^-  ) \  =  \ 0 \\ 
\]

The situation for $\stackrel{\circ}{\dlap} (E^+ - E^-  )$ is more subtle and requires detailed investigations. We point out that the computation of the limits 
involving $\mathbf{\Psi}$ and $\mathbf{\Psi'}$ and $\Omega'$ are crucial. \\ \\ 
In the EV case, Christodoulou and Klainerman prove the corresponding result in lemma 17.0.2, on
page 504 of \cite{sta}. Here, we establish the new results where 
the null fluid term $T_{\Lu \Lu} $ and its
limit $T_{\Lu \Lu}^*$ change the picture. 
Whereby it is used that 
\be \label{Dlap4null1}
\Dlap_4 T_{\Lu \Lu} = \Dlap_4 \mathcal{N} 
= - tr \chi \mathcal{N} + l.o.t. 
\ee
This is a direct consequence from (\ref{nablaLN1}) and results thereafter.

Calculating the limits
(\ref{limitsPsiOmega*1}), applying  
(\ref{Dlap4null1}), (\ref{trianglePsi1}) and (\ref{trianglePsip1}), we 
derive formulas for 
$\mathbf{\Psi}$, $\mathbf{\Psi'}$, $\Omega$, $\Omega'$. 
We find 
\beas
\Omega' & = & 
- \frac{1}{2^{\frac{3}{2}} 4 \pi} \int_{- \infty}^{+ \infty} \Big{ \{ } 
\int_{S^2} \frac{\mid \Xi \mid^2 (u', \omega' ) - \overline{\mid \Xi \mid^2 (u')}
}{( 1 - \omega \omega' )^{\frac{1}{2}}} 
d \omega' 
+ 
4 \pi 
\int_{S^2} \frac{T_{\Lu \Lu}^* (u', \omega' ) - \overline{T_{\Lu \Lu}^* (u')}
}{( 1 - \omega \omega' )^{\frac{1}{2}}} 
d \omega' 
\Big{ \} } 
d u'     \\ 
& & \ \ 
- \frac{1}{2} 
 \int_{- \infty}^{+ \infty} 
 \Big{ \{ } 
sgn(u - u') 
\Big( \Big(
\mid \Xi \mid^2 (u', \omega') - \overline{\mid \Xi \mid^2 (u')} \Big)
+ 4 \pi  \Big(T_{\Lu \Lu}^* (u', \omega') -  \overline{T_{\Lu \Lu}^* (u')} 
\Big)
\Big)
\Big{ \} } du' \\ 
\Omega & = & 
\frac{1}{2^{\frac{3}{2}} 4 \pi} \int_{- \infty}^{+ \infty} \Big{ \{ } 
\int_{S^2} \frac{\mid \Xi \mid^2 (u', \omega' ) }{( 1 - \omega \omega'
)^{\frac{1}{2}}} 
d \omega' 
+ 
4 \pi 
\int_{S^2} \frac{T_{\Lu \Lu}^* (u', \omega' ) }{( 1 - \omega \omega'
)^{\frac{1}{2}}} 
d \omega' 
\Big{ \} } 
d u'     \\ 
& & \ \ 
+ \frac{1}{2} 
 \int_{- \infty}^{+ \infty} 
 \Big{ \{ } 
sgn(u - u') 
\Big( 
\mid \Xi \mid^2 (u', \omega') + 4 \pi T_{\Lu \Lu}^* (u', \omega') 
\Big)
\Big{ \} } du' \\ 
\mathbf{\Psi'} & = & 
\frac{1}{2^{\frac{1}{2}} 4 \pi} \int_{- \infty}^{+ \infty} \Big{ \{ } 
\int_{S^2} \frac{\mid \Xi \mid^2 (u', \omega' ) - \overline{\mid \Xi \mid^2 (u')} 
}{( 1 - \omega \omega' )^{\frac{1}{2}}} 
d \omega' 
+ 
4 \pi 
\int_{S^2} \frac{T_{\Lu \Lu}^* (u', \omega' )  - \overline{T_{\Lu \Lu}^* (u')} 
}{( 1 - \omega \omega' )^{\frac{1}{2}}} 
d \omega' 
\Big{ \} } 
d u' \\ 
\mathbf{\Psi} & = & 
- \frac{1}{2^{\frac{1}{2}} 4 \pi} \int_{- \infty}^{+ \infty} \Big{ \{ } 
\int_{S^2} \frac{\mid \Xi \mid^2 (u', \omega' ) }{( 1 - \omega \omega'
)^{\frac{1}{2}}} 
d \omega' 
+
4 \pi 
\int_{S^2} \frac{T_{\Lu \Lu}^* (u', \omega' ) }{( 1 - \omega \omega'
)^{\frac{1}{2}}} 
d \omega' 
\Big{ \} } 
d u' \\ 
\eeas
By straightforward computations we find the following: Investigating the difference of the limits as 
$u \to + \infty$ and $u \to - \infty$ 
in (\ref{systeps*5}), there are no contributions from $\stackrel{\circ}{\slap} \Psi$, $\Psi'$. 
Only terms in
$\Omega'$ contribute. 
It follows directly that $\Omega'$ tends to limits $\Omega'^+ (\cdot)$ and $\Omega'^- (\cdot)$
as $t \to \infty$ and $t \to - \infty$, respectively. 
From this we derive 
\be
\Omega'^+ (\cdot) - \Omega'^- (\cdot) \ = \ 
\int_{- \infty}^{+ \infty} \big( \ 
\mid \Xi (u, \cdot) \mid^2 - \overline{\mid \Xi (u, \cdot) \mid^2} 
+ 4 \pi T_{\Lu \Lu}^* (u, \cdot) - 4 \pi  \overline{T_{\Lu \Lu}^* (u, \cdot)} 
\  \big) du \ \ . 
\ee
We conclude 
\bea
\stackrel{\circ}{\dlap} (E^+ - E^-  ) \ & = & \ - \Omega'^+  + \Omega'^-   \\ 
\ & = & \ 
\int_{- \infty}^{+ \infty} \big( \ -
\mid \Xi (u, \cdot) \mid^2 + \overline{\mid \Xi (u, \cdot) \mid^2} 
-  4 \pi T_{\Lu \Lu}^* (u, \cdot) + 4 \pi  \overline{T_{\Lu \Lu}^* (u, \cdot)} 
\  \big) du \ \ .  \nonumber 
\eea
This yields 
\be \label{phi**1}
(E^+ - E^- ) \ = \ \stackrel{\circ}{\nlap} \Phi 
\ee
where $\Phi$ is the solution of 
\[
\stackrel{\circ}{\slap} \Phi \ = \ - \Omega'^+  + \Omega'^-   \ \ \mbox{ on } S^2 \
\ ,
\]
with $\bar{\Phi} = 0$ on $S^2$. 

Now, we need equation (\ref{*circSigma**3}) from above 
\[
 \stackrel{\circ}{\dlap} \Sigma = E \ . 
\]

Relating result (\ref{phi**1}) to this equation, where in the latter we first take the limits as $u \to \infty$ and $u \to - \infty$, 
we conclude 
\be \label{*circSigma**1} 
\stackrel{\circ}{\dlap} (\Sigma^+ - \Sigma^-) = E^+ - E^-  =  \stackrel{\circ}{\nlap} \Phi  \ \ , 
\ee
which is equation (\ref{Thm*divSigma+-*2}).

This concludes the proof of the theorem. 
\subsection{Limit for $r$ as $t \to \infty$ on Null Hypersurface $C_u$}
We shall use the fact that the constraint on the spacelike scalar
curvature, which is given by 
\begin{equation*}
R=\left \vert k\right \vert ^{2}+R_{00},
\end{equation*}%
differs from the constraint in the vacuum case only by the term $R_{00}$.  \\ \\ 
We can now prove the following results. 
\begin{The} \label{r*1}
As $t \to \infty$ we obtain on any null hypersurface $C_u$ 
\[
r = t - 2 M(\infty) \log t + O(1)  \ \ . 
\]
\end{The}
{\bf Proof:} 
We recall from \cite{sta}, p. 503, with $\phi' = \phi -1$, 
\beas
\frac{dr}{dt} & = & \frac{r}{2} \overline{\phi tr \chi'} \\ 
& = &  
 \frac{r}{2} \overline{(1 + \phi')(\frac{2}{r} + (tr \chi' - \frac{2}{r}))} \\ 
& = & 
1 + \overline{\phi'} + O(r^{-2})  
\eeas 
In the last equality, we use equation (\ref{H2}).

Now, in the Einstein-null-fluid case, we have for $R_{00}$ the following expression in terms of the components $T_{LL}$ and $T_{\Lu \Lu}$ of the null fluid: 
\be \label{R00*1}
R_{00} = 8 \pi T_{00} = 2 \pi  ( T_{\Lu \Lu} - T_{LL} ) 
\ee
Moreover, 
the lapse equation in our situation is given by 
\be \label{lapset*1}
\triangle \phi = ( \mid k \mid^2 + R_{00} ) \phi \ \ . 
\ee
We integrate the lapse equation (\ref{lapset*1}) on $H_t$ in the interior of $S_{t,
u'}$ to obtain 
\[
\int_{S_{t, u}} \nabla_N \phi' = \int_{u_0}^u du' \int_{S_{t, u'}} a \phi ( \mid k
\mid^2  + R_{00} ) \ \ . 
\]
In view of (\ref{R00*1}) and the fact that all the terms on the right hand side of
(\ref{R00*1}) except $T_{\Lu \Lu}$ are of 
lower order, we estimate 
\beas
\int_{S_{t, u}} \nabla_N \phi'  & = &   \int_{u_0}^u du' \int_{S_{t, u'}} a \phi (
\mid k \mid^2  + 2 \pi T_{\Lu \Lu} )  + l.o.t.  \\ 
\eeas
We see that 
\[
\int_{S_{t, u'}}  a \phi ( \mid k \mid^2  + 2 \pi  T_{\Lu \Lu})  \to 
\int_{S^2} \mid \Xi \mid^2 + 2 \pi T_{\Lu \Lu}^* \ \ . 
\]
Consider the Bondi mass loss formula in theorem \ref{Bondimassloss*1}. Then, as
$t \to \infty$ we conclude 
\be \label{nablaNM**1}
\int_{S_{t,u}} \nabla_N \phi' - 8 \pi M(u) = O(r^{-1}) 
\ee
on each $C_u$. 
In view of $\phi'$ we compute: 
\beas
\overline{\phi'} & = & \frac{1}{4 \pi r^2} \int_{S_{t,u}} \phi' = - \frac{1}{4 \pi}
\int_{B} div (r^{-2}  \phi' N) \\ 
& = & 
\frac{1}{4 \pi}  \int_{B} \big( 
- \frac{1}{a(r(t,u'))^2} \overline{a tr \theta} N \phi' 
+ \frac{1}{(r(t,u'))^2} \phi' \underbrace{(div N)}_{= tr \theta} 
+ \frac{1}{(r(t,u'))^2} \nabla_N \phi' 
 \big) \\ 
 & = & 
- \frac{1}{4 \pi} \int_u^{\infty} \frac{1}{(r(t,u'))^2}  du' 
\big(
 \int_{S_{t, u'}} 
 a \nabla_N \phi' + (a tr \theta - \overline{a tr \theta}) \phi'  
  \big) \\ 
 & = & 
 - \frac{1}{4 \pi} \int_u^{\infty} \frac{1}{(r(t,u'))^2}  du' 
 \big( 
  \int_{S_{t, u'}}  \nabla_N \phi'  
 \big) 
 + O(r^{-2}) \ \ . 
\eeas
where $B$ denotes the exterior of $S_{t,u}$.
Therefore, from (\ref{nablaNM**1}) it follows on $C_u$ as $t \to \infty$, 
\[
\overline{\phi'} (t, u) = - 2 \int_u^{\infty}  \frac{1}{(r(t,u'))^2} M(u') du' +
O(r^{-2}) 
= - \frac{2}{r} M(\infty) + O(r^{-2}) \ \ . 
\] 
Thus, we obtain on any cone $C_u$ for $t \to \infty$, 
\be
\frac{dr}{dt} = 1 - \frac{2}{r} M(\infty) + O(r^{-2}) \ \ . 
\ee
Thus, the statement of our theorem follows, which closes the proof. \\ \\ 

\section{Gravitational Wave Experiments}
\label{wave}

In the previous chapters, we derived a contribution from neutrino radiation to the nonlinear Christodoulou memory effect of gravitational waves. This effect will show as a permanent displacement of test masses in a laser interferometer gravitational-wave detector. 
In this section, we show how the mathematical results relate to experiment. 
In a typical source of a neutrino burst such as core-collapse supernovae or binary neutron star mergers, over a timescale of tens of seconds a huge amount of energy is radiated away in form of neutrinos. In particular, in the case of a supernova, it is expected that approximately $99\%$ of the gravitational binding energy of the remnant in the process is converted into neutrinos. See Scholberg's article \cite{ksch} for a recent review. 
In such a process, gravitational waves are emitted and the wave package is traveling at the speed of light along the null hypersurfaces of our spacetime. We may think of doing the experiment at null infinity of the spacetime. 

Our findings are two-fold: First, we discuss the instantaneous displacements of test masses occurring while the package is moving through the experiment. Second, we investigate the permanent displacements of test masses after the gravitational wave train has passed, namely the nonlinear Christodoulou memory effect. 
We prove that the contribution from neutrino radiation described as a null fluid has only lower order \footnote{Here and in what follows, the word `order' refers to decay behavior of the exact solution, not to any approximations. That is, `higher order' means `less decay'. For details, see \cite{1lpst1}, \cite{lydia2}, \cite{zip2}.} contribution to the first effect, but contributes at the same highest order as the `purely geometrical' term to the nonlinear Christodoulou effect. 
The information about the null fluid part is `encoded' in $\Sigma^+ - \Sigma^-$ and described in theorem \ref{displ*1}. Precisely, this latter term governs the permanent displacement as we show at the end of the present chapter.

Now, we briefly review the setup of such a detector with three test masses. A detailed explanation is given in Christodoulou's pioneering paper \cite{chrmemory}, and a derivation in the Einstein-Maxwell case is given in the article by the first present author with Chen and Yau in \cite{1lpst1}. 

Let us think of the experiment having a reference test mass $m_0$ at the location of the beam splitter. Initially, masses $m_1$ and $m_2$ are at equal distances $d$ from $m_0$ forming a right angle there. In an Earth-based detector such as LIGO, the masses are suspended by pendulums and thus are free. If the observation is performed in space as in LISA, then the masses are free by nature. In fact, in the first case, for time scales much shorter than the period of the pendulums the motion of the masses 
in the horizontal plane can be considered free. 
By laser interferometry, the distance of $m_1$ and $m_2$ from the reference mass $m_0$ is measured. Whenever the light travel times between the masses differ, then 
we see a difference of phase of the laser light at $m_0$. 

The three masses move along geodesics in spacetime. We denote the geodesic for $m_0$ by $\Gamma_0$. Let $T$ be the future directed tangent vector field of $\Gamma_0$ of unit magnitude. Moreover, let $t$ denote the arc length along $\Gamma_0$. 
At $\Gamma_0(0)$ we choose an orthonormal frame $(e_1, e_2, e_3)$ for the spacelike, geodesic hyperplane $H_0$. For each $t$ denote by $H_t$ the spacelike, geodesic hyperplane through $\Gamma_0(t)$ orthogonal to $T$. 
We obtain the orthonormal frame field $(T, e_1, e_2, e_3)$ along $\Gamma_0$ by parallel propagation of $(e_1, e_2, e_3)$. The latter being an orthonormal frame for every $H_t$ at $\Gamma_0(t)$. 
To a point in spacetime close to $\Gamma_0$ and lying in $H_t$ we can now assign cylindrical normal coordinates $(t, x^1, x^2, x^3)$. 

Assume that the source of the waves is in the $e_3$-direction, and that the light travel time corresponding to the distance $d$ is significantly shorter than the time scale of large variations of the spacetime curvature. Then the geodesic equation for the trajectories of $m_1$ and $m_2$ are replaced by the Jacobi equation (\ref{Jacobi**1}) measuring the geodesic deviation from $\Gamma_0$. With $R_{kTlT}  =  R  (e_k, T, e_l, T)$ it is 
\be \label{Jacobi**1}
\frac{d^2 x^k}{d t^2} \ = \ - \ R_{kTlT} \ x^l \ . 
\ee
The acceleration in (\ref{Jacobi**1}) is controlled by the curvature $R_{kTlT}$. In order to reveal the roles played by the null fluid and by the gravitational part, we have to 
investigate the structure of $R_{kTlT}$. Thus, we decompose the latter into its Weyl and Ricci parts: 
\be \label{RiemWeylRic1}
R_{kTlT} = W_{kTlT} + \frac{1}{2} ( g_{TT} R_{kl} +  g_{kl} R_{TT} - g_{Tl} R_{kT} - g_{Tk} R_{lT} )  . 
\ee
The Einstein-null-fluid equations (\ref{Einsteinnullfluid1}) tell us that 
\[
R_{TT} = 8 \pi T_{TT} \ \ , 
\]
ensuring the following identity: 
\be
R_{TT} = 8 \pi T_{TT} = 2 \pi  ( T_{\Lu \Lu} - T_{LL})  . 
\ee
The worst decay behavior on the right hand side of (\ref{RiemWeylRic1}) occurs in $R_{TT}$, namely we find it in the null fluid component 
$T_{\Lu \Lu}$. 

To take limits at null infinity, we change to the null frame with $L = T - e_3$ and $\underline{L} = T + e_3$. 
Then the leading components of the curvature can be expressed as 
\beas
\underline{\alpha}_{AB} & = &  R(e_A, \underline{L}, e_B, \underline{L}) \\ 
\underline{\alpha}_{AB} & = & \frac{A_{AB}}{r} + o  (r^{-2}) \ , 
\eeas
and the leading component of the null fluid as
\bea
T_{\Lu \Lu}  \ & = & \ \frac{ T_{\Lu \Lu}^* }{r^2} \ + \ l.o.t. \ . 
\eea

We now observe that the null fluid enters the right hand side of the Jacobi equation at order $(r^{-2})$. 
As a consequence the null fluid does not contribute at leading
order to the deviation measured. 
This brings us back to the situation for the Einstein vacuum equations investigated by Christodoulou in \cite{chrmemory}. 
At leading order, our result coincides with his 
\be \label{Jac***4}
\frac{d^2 \ x^k_{\ (A)}}{d \ t^2} \ =  \ - \ \frac{1}{4} \ r^{-1}  \ A_{AB} \ x^l_{\
(B)} \ + \ 
O \ (r^{-2}) 
\ee
Similarly to \cite{chrmemory} and \cite{1lpst1} we find that in the Einstein-null-fluid case 
there is no acceleration in the vertical direction to leading order $(r^{-1})$. 
Before the wave package travels through the experiment, the masses $m_1$ and $m_2$ are at rest at equal distance $d$ and at right angles from 
$m_0$. That is, we have the initial conditions as $t \to - \infty$: \\ 
$x^3_{\ (A)}  =  0 \ , \ \dot{x}^3_{\ (A)}  = 0 \ , \ 
x^B_{\ (A)} =  d  \delta^B_A  \ , \ 
\dot{x}^B_{\ (A)}  =  0$. 
As the right hand side is very small, one can substitute the initial values on the
right hand side. 
Then the motion is confined to the horizontal plane. To leading order it is: 
\be
\stackrel{\cdot \cdot}{x}^A_{\ (B)} \ = \ - \ \frac{1}{4} \ r^{-1} \ d \ A_{AB}  \
\ . 
\ee
Integrating yields 
\be 
\dot{x}^A_{\ (B)} \ (t) \ = \ - \ \frac{1}{4} \  d \  r^{-1} \ 
\int_{- \infty}^t \ A_{AB} \ (u) \ d u \ . 
\ee
To derive the new result for the Einstein-null-fluid equations, we revisit formula (\ref{reschrmem****4}) in 
\cite{chrmemory}. 
At this point, the identities (\ref{Xiu*1}) and (\ref{Sigmau*1}) will be applied. First, using equation (\ref{Xiu*1}), namely 
$
\frac{\partial \Xi }{\partial u}  = -  \frac{1}{4} \ A $
with 
$\lim_{\mid u \mid \to \infty}  \Xi  =  0$ 
we have
\be
\Xi \ (t) \ = \ 
- \ \int_{- \infty}^t \ A_{AB} \ (u) \ d u 
\ee
and thus 
\be
\dot{x}^A_{\ (B)} \ (t) \ = \ \frac{d}{r}  \ \Xi_{AB} \ (t) \ . 
\ee
As $\Xi \to 0$ for $u \to \infty$, the test masses return to rest after 
the passage of the gravitational wave train. 
Next, using (\ref{Sigmau*1}), that is $\frac{\partial \Sigma}{\partial u } =  -  \Xi$, 
another integration yields
\be
x^A_{\ (B)} \ (t) \ = \ - \ (\frac{d}{r}) \ (\Sigma_{AB} \ (t) \ - \ \Sigma^-) \ . 
\ee
Now, we take the limit $t \to \infty$ to obtain 
\be \label{reschrmem****4}
\triangle \ x^A_{\ (B)} \ = \ - \ (\frac{d}{r}) \ (\Sigma^+_{AB} \ - \
\Sigma^-_{AB}) \ . 
\ee
Thus, we find that the test masses are permanently displaced. In particular, 
$\Sigma^+ - \Sigma^-$ is equivalent to an overall displacement of the test masses. Precisely this term, in theorem \ref{displ*1}, is proven 
to exhibit a contribution from the null fluid besides the purely gravitational part. 

Thus, we find that the instantaneous displacements of test masses are not affected at highest order by the null fluid. However, the null fluid does contribute at highest order to the permanent displacement of the masses and therefore enlarges the nonlinear Christodoulou memory effect of gravitational waves. \\

{\bf Acknowledgment:} We thank Demetrios Christodoulou and Bob Wald for discussions. 
L. Bieri is supported by NSF grants DMS-1253149 and DMS-0904760.  D. Garfinkle is supported by NSF grants PHY-0855532 and PHY-1205202.

\vspace{2cm}


 
%
\vspace{10pt}
{\scshape Lydia Bieri \\ 
Department of Mathematics \\ 
University of Michigan \\ 
Ann Arbor, MI 48109, USA} \\ 
lbieri@umich.edu \\ \\ 
{\scshape David Garfinkle \\ 
Department of Physics \\ 
Oakland University \\ 
Rochester, MI 48309, USA \\
and Michigan Center for Theoretical Physics \\
Randall Laboratory of Physics\\
University of Michigan\\
Ann Arbor, MI 48109, USA} \\ 
garfinkl@oakland.edu

\end{document}